\documentclass[preprint]{aastex631}
\usepackage{natbib}
\accepted{25 June 2025}

\submitjournal{AJ}

\begin{document}

                                                                                                                                              
\title{Bringing the Norma Dark Cloud to Light in X-rays}

\correspondingauthor{Stephen L. Skinner}
\email{stephen.skinner@colorado.edu}

\author{Stephen L. Skinner}
\affiliation{Center for Astrophysics and
Space Astronomy (CASA), Univ. of Colorado,
Boulder, CO, USA 80309-0389}

\author{Manuel  G\"{u}del}
\affiliation{Dept. of Astrophysics, Univ. of Vienna,
T\"{u}rkenschanzstr. 17,  A-1180 Vienna, Austria}

\author{Luisa M. Rebull}
\affiliation{Infrared Science Archive (IRSA), IPAC, 1200 E. California Blvd., 
California Institute of Technology, Pasadena, CA USA 91125}


                                                                                                                                              
\newcommand{\ltsimeq}{\raisebox{-0.6ex}{$\,\stackrel{\raisebox{-.2ex}%
{$\textstyle<$}}{\sim}\,$}}
\newcommand{\gtsimeq}{\raisebox{-0.6ex}{$\,\stackrel{\raisebox{-.2ex}%
{$\textstyle>$}}{\sim}\,$}}
\begin{abstract}
\small{
 The filamentary dark cloud complex in Norma reveals signs of active low-mass star 
 formation including protostars, H$\alpha$ emission line  stars, 
 Herbig Haro objects, and the eruptive FU Orionis-like star V346 Nor.
 We present results of the first pointed X-ray observations of the Norma dark cloud,
 focusing on the westernmost Sandqvist 187 region.  Chandra detected 75 X-ray
 sources and a complementary XMM-Newton observation detected 92 sources within the
 Chandra field-of-view, of which 46 are cross-matched to Chandra,
 yielding  121 unique X-ray sources. We present a catalog of X-ray sources 
 along with basic X-ray properties and candidate IR and optical counterparts.
 Existing near-IR photometry reveals several X-ray sources with color excesses 
 as typical of young stars with disks. Gaia parallaxes 
 single out foreground stars and X-ray sources with distances of 500 - 1000 pc 
 that are probable cloud members. The known emission line stars Sz 136 and Sz 137 
 were detected but V346 Nor was not.  
 Interestingly, the optical and IR counterparts of the brightest Chandra source
 are not known with certainty but the prime suspects are very faint.
 Thus, the nature of the  object responsible for the bright X-ray  
 emission remains speculative.
 The X-ray observations presented here will serve as a
 pathfinder for identifying and characterizing the young stellar population
 in the Norma dark cloud.
}
\end{abstract}



\section{Introduction}
\noindent 
Filamentary dark clouds are
identified as elongated regions of dense dust and gas silhoutted against
the field of background stars. They have acquired considerable interest
as sites of active star formation and for their elongated shape, which
bears little resemblance to the spherical clouds historically assumed
in cloud-collapse star-formation models. Previous studies, most notably
with {\em Herschel} (Andr\'{e} 2017), dramatically reveal
the ubiquity of filamentary structures in interstellar clouds.
The densest filaments contain prestellar cores and trace the
birth sites of young stars. The
physical mechanisms that give rise to the filamentary structures and
chained core formation are an active  research area and several factors 
may be involved including magnetic fields, shocks,  
turbulence, and  gravitational instabilities (e.g. Andr\'{e} 2017; Myers 2017).

The Norma dark cloud is composed of two elongated filamentary
structures Sa 187 and Sa 188 (Sandqvist 1977).
The two filaments span $\approx$40$'$ EW $\times$ 15$'$ NS 
(Reipurth \& Nielbock 2008, hereafter RN08). 
The cloud is undergoing active low-mass star formation as reviewed by RN08.
The western filament Sa 187 discussed herein harbors a variety of
young stellar objects (YSOs) including the eruptive star V346 Nor
which drives HH 57, the H$\alpha$ emission line stars
Sz 136 and Sz 137 (Schwartz 1977), a small
reflection nebula Re 13 that conceals a newborn star
thought to drive HH 56 (Prusti et al. 1993; RN08),
and several compact millimeter cores 
(Nielbock \& Chini 2005, hereafter NC05).

Estimates of the cloud distance span a wide range as
summarized by RN08. The near side of the cloud is estimated 
to lie at d = 440$\pm$50 pc (NC05). The
distance to the far side is rather
uncertain but Graham \& Frogel (1985) estimated
an upper limit of 1100 pc. More recent Gaia DR3
data show that it extends to at least $\sim$900 pc
as determined from parallax distances of the cloud members 
Sz 136 and Sz 137. 
Previous studies have usually assumed an
average cloud distance of 700 pc based on Graham \& Frogel (1985)
which we will adopt here unless more specific Gaia distances
are available. The extinction is variable across
the cloud with an average A$_{\rm V}$ = 7.6 mag (NC05).

Despite multiple signposts of active star formation,
the Norma dark cloud
has not previously been observed in X-ray pointed observations
which motivated the new observations  presented here.
The X-ray observations
discussed below are centered near the star V346 Nor located
in Sa 187 in the western part of the cloud complex, as shown in
Figure 1 and the wide-field image of RN08.
We provide a catalog of X-ray sources in the Sa 187 region
based on {\em Chandra} observations along with basic X-ray properties
and candidate IR counterparts. Complementary XMM-Newton
X-ray data are used to clarify and extend the Chandra results.

\section{X-ray Observations and Data Reduction}

\subsection{Chandra}

Chandra obtained four observations of the Norma Dark Cloud in 2022
(Cycle 22)
using the Advanced CCD Imaging Spectrometer (ACIS-I). 
Pointing was centered near V346 Nor at J163232.2$-$445530.7
and Galactic coordinates ($l$=338.55$^{\circ}$, $b$=$+$2.12$^{\circ}$).
Livetimes, which exclude overhead such as readout times for 
each 3.2 s exposure, were
13.777 ks (ObsId 23387, 28 Jan.), 
15.912 ks (ObsId 26283, 29 Jan.),
14.773 ks (ObsId 24511, 4 Oct.), and
14.770 ks (ObsId 27464, 4-5 Oct.).
The merged observations have a total livetime of 59.232 ks. 
ACIS-I consists of four CCDs giving a combined 
16.9$'$ $\times$ 16.9$'$ field-of-view (FoV).
The energy coverage is $\approx$0.3 - 10 keV
but the ACIS-I effective area (A$_{\rm eff}$) below 0.5 keV is 
low\footnote{For more information on ACIS-I see 
https://cxc.harvard.edu/cal/acis/ .}. 
The ACIS spatial resolution at 1 keV for near on-axis sources is
$\approx$0$''$.7 (80\% encircled power radius) but increases with
off-axis angle. 
The absolute accuracy of Chandra ACIS-I X-ray positions 
relative to accurately known optical positions for sources
within 3$'$ of the aimpoint and sim-Z at the nominal ACIS-I detector value
has a radius of 0$''$.93 (90\% confidence)\footnote{Section 5 of the 
Chandra Proposer's Observatory Guide:~ 
https://cxc.harvard.edu/proposer/POG}.
The X-ray position uncertainty is larger for sources further from the aimpoint and
depends on source count statistics with larger uncertainties for fainter sources.

Data were reduced using Chandra Interactive Analysis
of Observations (CIAO vers. 4.14) software with
CALDB v 4.9.8 calibration data\footnote{For more information on
CIAO and CALDB see https://cxc.cfa.harvard.edu.}. 
X-ray source detection was carried out using the CIAO wavelet 
detection tool {\em wavdetect} on energy-filtered (0.5 - 7 keV) images 
for each observation and for the merged observations. The four 
observations were merged using  CIAO {\em merge\_obs}.
Several options for combining the per-observation point-spread 
function (PSF) maps are provided by {\em merge\_obs} and
we have used the psfmerge=exptime option which weights each pixel
by the exposure time of the observation, as described in the 
{\em merge\_obs} 
documentation\footnote{https://cxc.cfa.harvard.edu/ciao/ahelp/merge\_obs.html\#plist.psfmerge}.
Wavelet scales of 2, 4, and 8 native pixels were used in {\em wavdetect}, 
with a native  ACIS pixel size = 0.$''$49  in unbinned images. 
At these scales, {\em wavdetect} is sensitive to point
sources and sources with modest spatial extension, where the latter can
occur for far off-axis sources due  to PSF broadening. The detection threshold 
in {\em wavdetect} was set to sigthresh=10$^{-7}$ as
appropriate for the merged image size 
and results in no more than $\approx$2-3 false sources 
arising from strong background fluctuations\footnote{For 
further details on {\em wavdetect} see 
https://cxc.cfa.harvard.edu/ciao/ahelp/wavdetect.html.}. 
In fact, background was low and averaged
0.052 counts/arcsec$^{2}$ (0.5- 7 keV), 
or 0.0125 counts per native pixel in the merged image.

The merged source list in Table 1 contains 75 Chandra sources
as plotted in Figure 1. Table 1 also includes
basic X-ray properties and candidate IR and optical identifications. 
The source list includes 46 sources that were detected in 
one or more of the Chandra observations, 21 sources 
detected in two or more, and
29 faint sources that were only detected in the merged observations.

The  median photon energy (E$_{50}$) and hardness ratio 
H.R. = cts(2-7 keV)/cts(0.5-7 keV) for
each source were computed using events inside 
{\em wavdetect} 3$\sigma$ source position error ellipses
and were background subtracted
using the method described in Appendix C
of Hong, Schlegel, \& Grindlay (2004).
To assess variability within a single observation
CIAO {\em glvary} was executed separately for each 
observation in which the source was detected. 
We did not attempt to assess variability of the faintest
sources that were only detected in the merged data.
For those sources detected in two or more observations
we compared the net count rates for the different 
observations. If the rates in two or more observations
differed by at least 3$\sigma$ the source was deemed variable.

Observation-specific spectra and response  files were extracted for
some sources of interest using CIAO {\em specextract}.
Spectra were fitted using  models in XSPEC  vers. 12.10.1
to determine basic parameters such as equivalent neutral hydrogen
absorption column density (N$_{\rm H}$), plasma temperature in 
energy units (kT), and X-ray flux (F$_{x}$).
However, most sources lack sufficient counts in a single observation
to extract and reliably model observation-specific spectra. 
Although spectra extracted from merged events would contain more
counts, they are not suitable for analysis since the merged events
lack sufficient information to generate correct response files.

To estimate the on-axis detection limit in the merged 
Chandra observations we consider
the faintest sources which contain 7$\pm$3 net counts, or as few as
4 net counts allowing for the uncertainty. The mean count rate
needed to accumulate 4 net counts in 59.232 ks is 0.067 c ks$^{-1}$.
We assume an absorbed  thermal plasma spectrum with a temperature range of 
kT = 1 - 3 keV, as typical for a T Tauri star (TTS). 
Adopting a mean cloud extinction A$_{\rm V}$ = 7.6 mag gives
N$_{\rm H}$ = 1.44 $\times$ 10$^{22}$ cm$^{-2}$ using a conversion 
N$_{\rm H}$ (cm$^{-2}$) = 1.9$\pm$0.3 $\times$ 10$^{21}$$\cdot$A$_{\rm V}$ (mag)
(Gorenstein 1975; Vuong et al. 2003). 
Chandra Cycle 22 PIMMS simulations 
using the above input parameters give absorbed and unabsorbed flux
detection limits (0.5-7 keV) of 
F$_{x,abs}$ = (0.7 - 1.0) and F$_{x,unabs}$ = (1.8 - 4.1)
$\times$10$^{-15}$ ergs cm$^{-2}$ s$^{-1}$, 
where the low and high values in parentheses correspond
to kT = 1 keV and kT = 3 keV respectively. 
At a distance d$_{\rm pc}$ in parsecs the intrinsic (unabsorbed) X-ray luminosity
limit is L$_{x}$(0.5-7 keV) = (2.1 - 4.9)$\times$10$^{23}$d$_{\rm pc}^{2}$ ergs s$^{-1}$.
At the mean cloud distance d = 700 pc the limit is
L$_{x}$(0.5-7 keV) = (1.0 - 2.4)$\times$10$^{29}$ ergs s$^{-1}$.
This is sufficient to detect subsolar mass TTS
(Preibisch et al. 2005; Telleschi et al. 2007a).
The detection limit depends rather sensitively on the 
X-ray absorption N$_{\rm H}$, especially for soft sources,
being higher for more heavily absorbed sources, all other
factors being equal. The position of the source on the 
detector also affects the detection limit since the 
ACIS-I effective area decreases with increasing off-axis
angle as described in the Chandra Proposer's Observatory Guide 
(POG)\footnote{https://cxc.harvard.edu/proposer/POG/html/chap6.html\#chap:acis}. 

To compare with the above detection limit based on PIMMS we have
computed F$_{x,abs}$(0.5-7 keV) for Chandra source CXO 32 (Table 1) using 
the method described by Broos et al. (2010). This source was only
detected in the merged observations with 7$\pm$3 net counts 
and a median photon energy E$_{50}$ = 4.24 keV so lacks sufficient 
counts for spectral extraction and flux measurement.
The absorbed flux in units of 10$^{-14}$ ergs cm$^{-2}$ s$^{-1}$ is
F$_{x.abs}$ = (1.602 $\times$ 10$^{-12}$$\cdot$E$_{50}$$\cdot$Rate)/A$_{\rm eff}$
where E$_{50}$ is in keV, net Rate is in c ks$^{-1}$, and the ACIS-I
effective area A$_{\rm eff}$ is in units of cm$^{2}$. In the formulation of
Broos et al. (2010) the average value of A$_{\rm eff}$ over the specified
energy range is used.  Since CXO 32 is offset
by only 0$'$.13 from the aimpoint we use the on-axis ACIS-I 
A$_{\rm eff}$ values for Cycle 22\footnote{https://cxc.harvard.edu/proposer/}
evaluated on an equally spaced energy grid with a step size of 0.1 keV.
This yields A$_{\rm eff,avg}$(0.5-7 keV) = 264 cm$^{2}$. 
For CXO 32 the net Rate = 0.13$\pm$0.05  c ks$^{-1}$ gives 
F$_{x,abs}$(0.5-7 keV) = 3.3$\pm$1.4 $\times$ 10$^{-15}$  ergs cm$^{-2}$ s$^{-1}$,
a few times greater than the above PIMMS detection limit. If we use
the higher value A$_{\rm eff}$ = 369 cm$^{2}$ evaluated at E$_{50}$ = 4.24 keV
instead of the average over the full energy range the computed value 
of F$_{x,abs}$ decreases by a factor of 0.7.

\subsection{XMM-Newton}
We make use of a companion 
XMM-Newton observation to clarify and extend Chandra results.
Chandra's higher spatial resolution is a clear advantage when 
observing crowded star-forming regions near the Galactic plane.
But XMM-Newton has higher A$_{\rm eff}$, especially below 1 keV,
and provides more spectral counts in high-interest 
sources, thus yielding more reliable spectral fit parameters
for sources that were only faintly detected by Chandra.

XMM-Newton observed the Norma Dark Cloud in a single
long exposure of $\approx$77 ks on 28 Feb. - 1 Mar. 2021, 
about 11 months prior to the first Chandra observation.
The primary instrument was the
European Photon Imaging Camera (EPIC) which provides
CCD imaging spectroscopy in the $\approx$0.1 - 12 keV energy range
from the pn camera (Str\"{u}der et al. 2001)
and two nearly identical MOS cameras (MOS1 and  MOS2; Turner et al. 2001).
The EPIC FoV spans $\approx$30$'$ and 
the angular resolution is $\approx$6$''$ (PSF FWHM).
EPIC relative position astrometric accuracy after applying 
spacecraft pointing corrections is $\approx$1$''$.5 (Kirsch et al. 2004). 

The X-ray source list provided from the Science Operations 
Center (SOC) pipeline processing was used to identify XMM-Newton 
sources captured within the Chandra ACIS-I field of view. 
This spatial masking resulted in the 92 sources listed in Table 2.
The positions of these 92 sources were compared with those 
of the 75 Chandra sources by overlaying all positions on the
ACIS-I and EPIC images. This revealed 46 sources whose position uncertainty
regions overlapped and they are identified as matched sources in Tables 1 and 2.

The Digital Sky Survey (DSS) POSS2 image overlay in 
Figure 1 shows the positions
of the 92 XMM-Newton sources detected within the Chandra FoV.
Many of the XMM-Newton sources
detected near the periphery of the ACIS-I field were undetected
by Chandra. Chandra's sensitivity to point sources is lower at such far
off-axis angles and full spatial overlap of the four observations
was not obtained at the periphery due to differences in roll angle
and ACIS-I footprint, resulting in lower exposure times at the periphery.

Table 2 also includes net (background subtracted) pn and MOS1$+$2 counts 
for each source from the pipeline processing, hardness ratios H.R.(pn) computed  
from net pn count rate ratios using the standard
pipline processing energy bands (Bands 1-8) spanning the 0.2-12 keV range,
and candidate counterparts. 
The faintest XMM-Newton pn source 
is XMM 48 with 30$\pm$12 net pn counts and a 
broad-band flux from the pipeline processing
F$_{x,abs}$(0.2-12 keV) = 1.66 $\times$ 10$^{-15}$ ergs cm$^{-2}$ s$^{-1}$.

In order to perform more detailed analysis of  
sources of interest the pipeline data were further reduced
using the XMM-Newton Science Analysis System software
(SAS vers. 20.0)\footnote{https://xmm-tools.cosmos.esa.int/external/xmm\_user\_support/documentation/sas\_usg/USG/} 
with recent calibration data.
Custom time filtering of the data for each instrument
was applied to remove intervals of high background radiation
using procedures provided in the SAS documentation.
Energy-filtered light curves from source-free background regions 
were used to  identify time intervals of steady low-background 
emission. These good time intervals (GTIs) were then applied
to the pipeline event lists using SAS  {\em evselect}.
The time-filtered data yielded GTI livetimes of 
$\approx$46.4 ks (pn), $\approx$49.8 ks (MOS1),
and $\approx$51.6 ks (MOS2). 
X-ray spectra and associated spectral response files were extracted 
from the time-filtered events for selected sources using SAS {\em xmmselect}.
A circular region of radius = 20$''$ centered on the source was
used for spectral extraction, corresponding to
$\approx$77\% (pn) and $\approx$74\% (MOS) encircled energy.
A smaller extraction region was used for the star Sz 136 to
mitigate contamination from a nearby source (Sec. 4.4).
Background spectra were extracted from source-free regions near the source.
Spectra were analyzed using XSPEC vers. 12.10.1, as for the Chandra spectra. 
We have not undertaken a detailed variability analysis of the XMM-Newton detections. 
The pipeline processing does not provide variability statistics
and the relatively high EPIC background can be misconstrued as
variability in fainter sources. Chandra's low ACIS-I background is 
preferred for assessing X-ray variability. 

\clearpage
\begin{figure}
\figurenum{1}
\includegraphics*[width=8.0cm,angle=0]{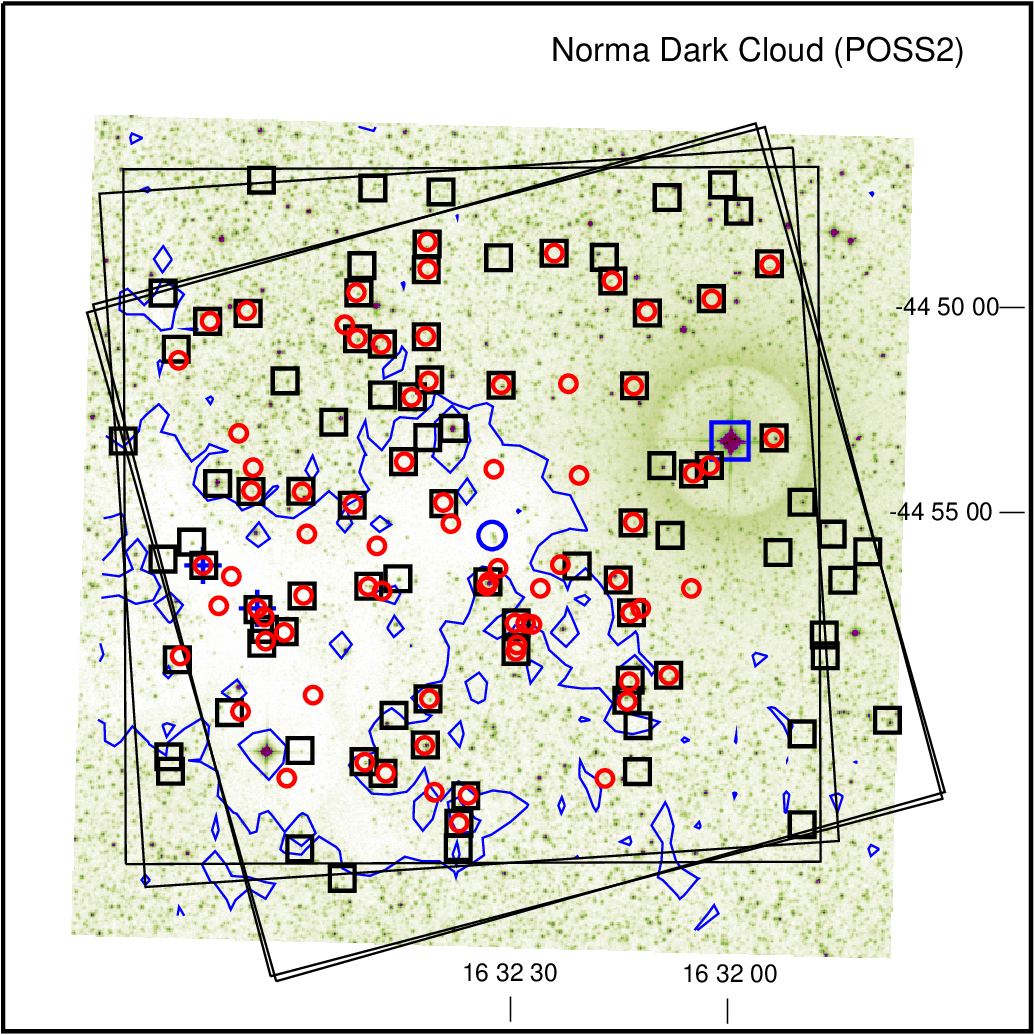}
\hspace*{0.6cm}
\includegraphics*[width=8.0cm,angle=0]{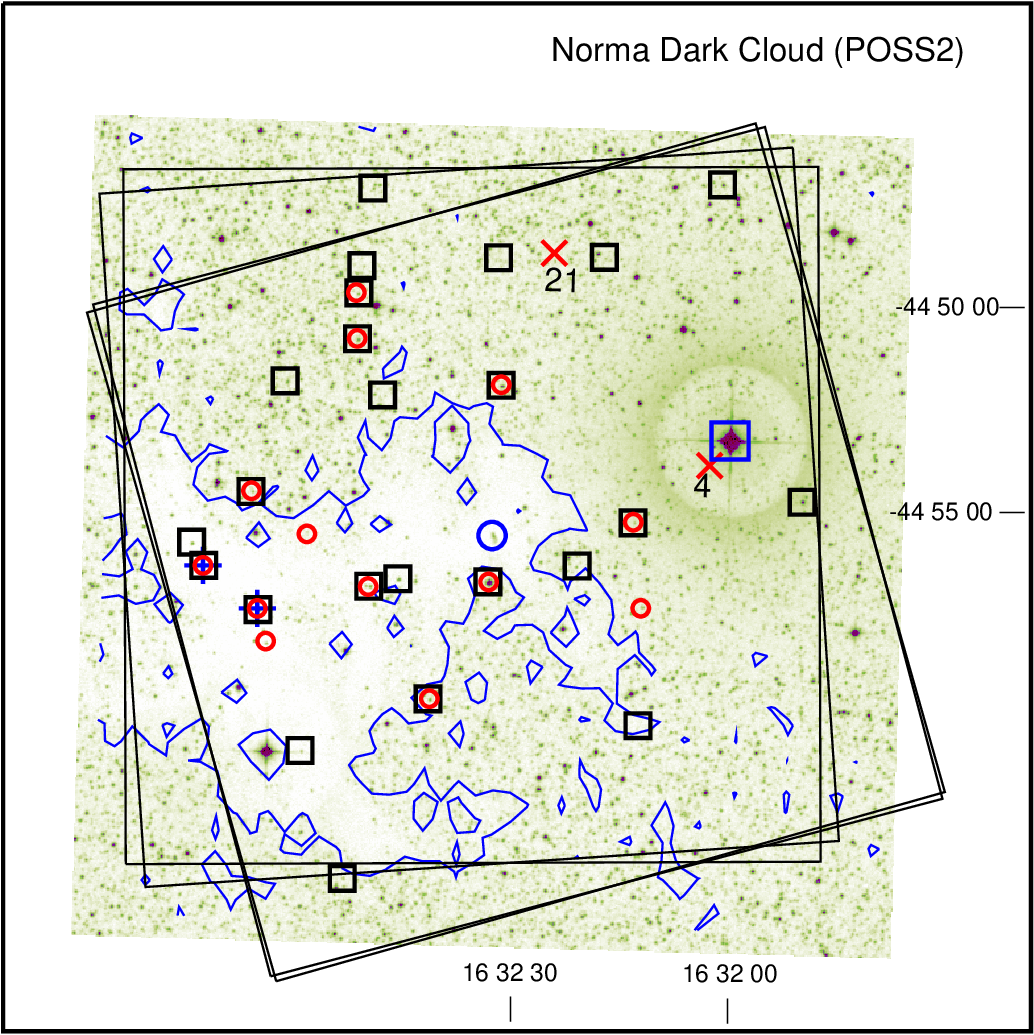}
\caption{
Left:~DSS POSS2 near-IR (0.85 $\mu$m) image of the Norma Dark Cloud
overlaid with Chandra X-ray sources (red circles, Table 1) and XMM-Newton sources
(black boxes, Table 2) captured on ACIS-I. 
The ACIS-I field-of-view for each of
the four observations at their respective roll angles is shown. 
The blue circle at center shows the position of the FUor-like star
V346 Nor (J163232.19$-$445530.7) near the aimpoint, and the blue box
marks the position of the IR-bright M1 III star V380 Nor, both undetected in X-rays. 
The two blue crosses near the left edge of the field mark the positions of 
the H$\alpha$ emission line stars Sz 136 (CXO 64) and Sz 137 (CXO 73).
The contour roughly outlines the darkest part of the cloud filament
and X-ray sources projected onto this dark region are identified in
Tables 1 and 2 as on-cloud (o) sources. \\
Right: Same as at left but showing only those X-ray sources having a 
Gaia DR3 match within 3$''$ and a parallax distance of 500 - 1000 pc.
These are identified as in-cloud (i) sources in Tables 1 and 2.
The two X points at (upper) right mark the positions of the bright X-ray
sources CXO 4 and CXO 21.
}
\end{figure}

\section{Results}

\subsection{General X-ray Properties}

Chandra source X-ray properties are summarized in Figure 2.
Most sources are faint with a median of 21 net counts.
A non-variable source would thus give an average of $\approx$5 net
counts in each of the four observations, insufficient
for analysis of observation-specific spectra.

All Chandra sources have E$_{50}$ $>$ 1 keV and the median value  
is E$_{50}$ = 2.88 keV. The median is quite high compared to
field stars viewed through low extinction  
and young stars in low-extinction star-forming regions (Wright et al. 2010a, 2023). 
The median hardness ratio is H.R. = 0.71.
For comparison, an isothermal $apec$ plasma spectrum from a stellar
X-ray source at kT = 1 - 4 keV viewed through the mean cloud
extinction A$_{\rm V}$ $\approx$ 7.6 mag (N$_{\rm H}$ = 1.44 $\times$ 10$^{22}$ cm$^{-2}$)
is predicted by Chandra Cycle 22 PIMMS simulations to have  
H.R. = 0.30  - 0.65 where the lower and upper values 
correspond to kT = 1 keV and 4 keV, respectively. 
Although the above high median value H.R. = 0.71 
is subject to the low ACIS sensitivity below 1 keV the
PIMMS simulations suggest that many Chandra sources are viewed 
through higher extinction than the cloud average.
These objects could be YSOs in denser parts of the cloud or heavily-obscured 
background sources. 
A clear exception is CXO 31, the low-extinction A2e 
star HD 328329 (Sec. 4.5).
Three faint sources are very hard with no counts below 2 keV and
H.R. = 1 (CXO 20,22,39). 
There is no obvious correlation between net count rate and H.R.
but brighter sources with rates $\geq$1 c ks$^{-1}$ tend to have high 
H.R. $\geq$ 0.6.

The median hardness ratio H.R.(pn) = 0.53 for the XMM sources in Table 2 
is somewhat less than the Chandra source median H.R. = 0.71.
One factor contributing to this difference is the better sensitivity
of EPIC pn below 1 keV. Nevertheless, very soft and 
hard sources clearly stand out in both the Chandra and 
XMM-Newton source lists, e.g. CXO 31 = XMM 39.

\subsection{Variability}

X-ray variability is commonly detected in low-mass pre-main sequence (PMS)
stars (e.g. T Tauri stars).
Five Chandra sources showed significant count rate variations within a single 
observation (CXO 1,6,38,68,73) but CXO 6 is faint (11$\pm$4 cts) 
and provides meager data to assess variability.
Comparison of count rates across the different
observations revealed variability for the brightest source (CXO 4),
CXO 64 (Sz 136), and CXO 73 (Sz 137). The second brightest
Chandra source CXO 21 may also be variable. These sources are 
discussed further below (Sec. 4).

\begin{figure}
\figurenum{2}
\includegraphics*[width=6.0cm,angle=-90]{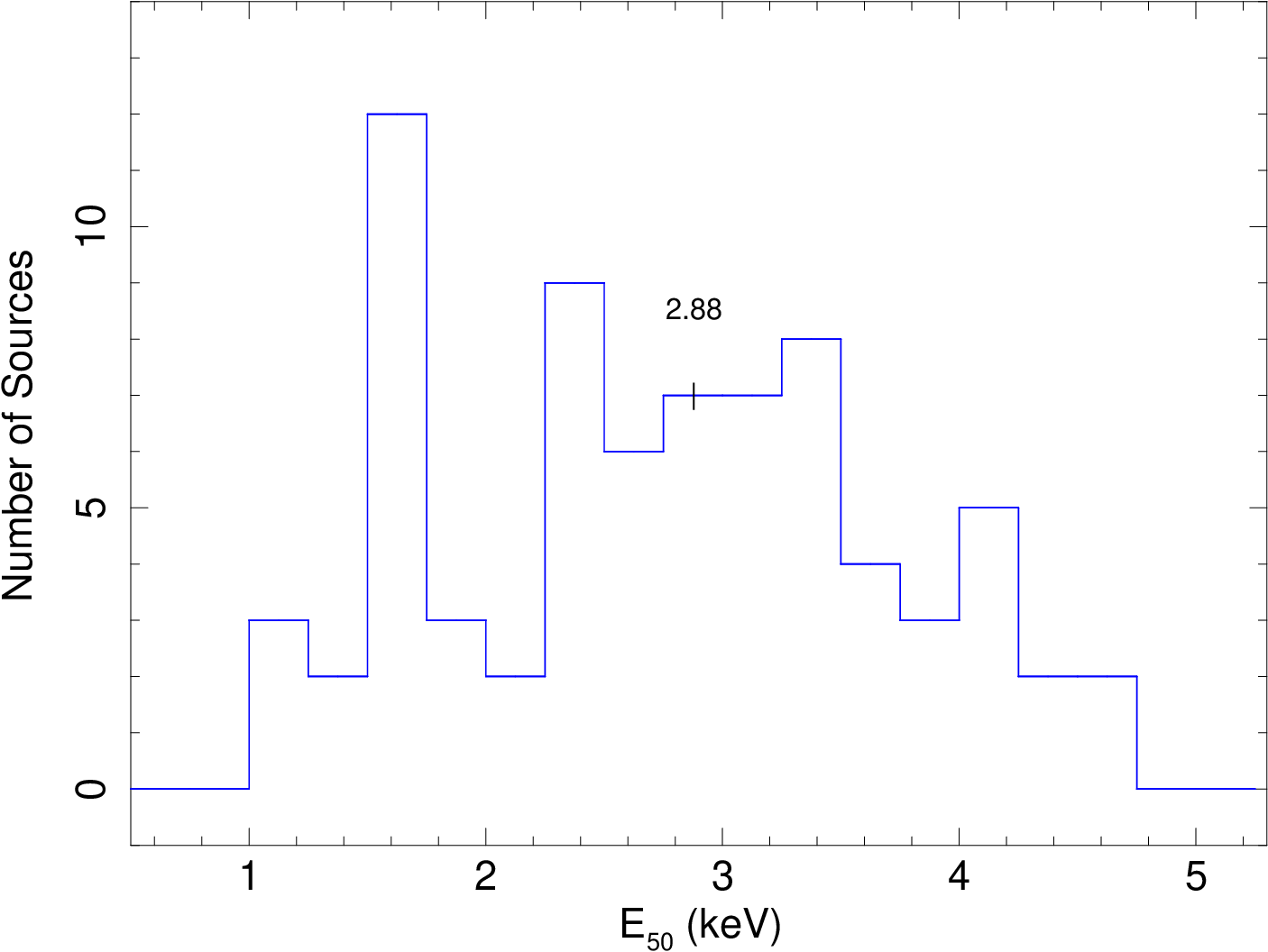} \\ \\
\vspace{0.3cm}
\vspace{0.3cm}
\includegraphics*[width=6.0cm,angle=-90]{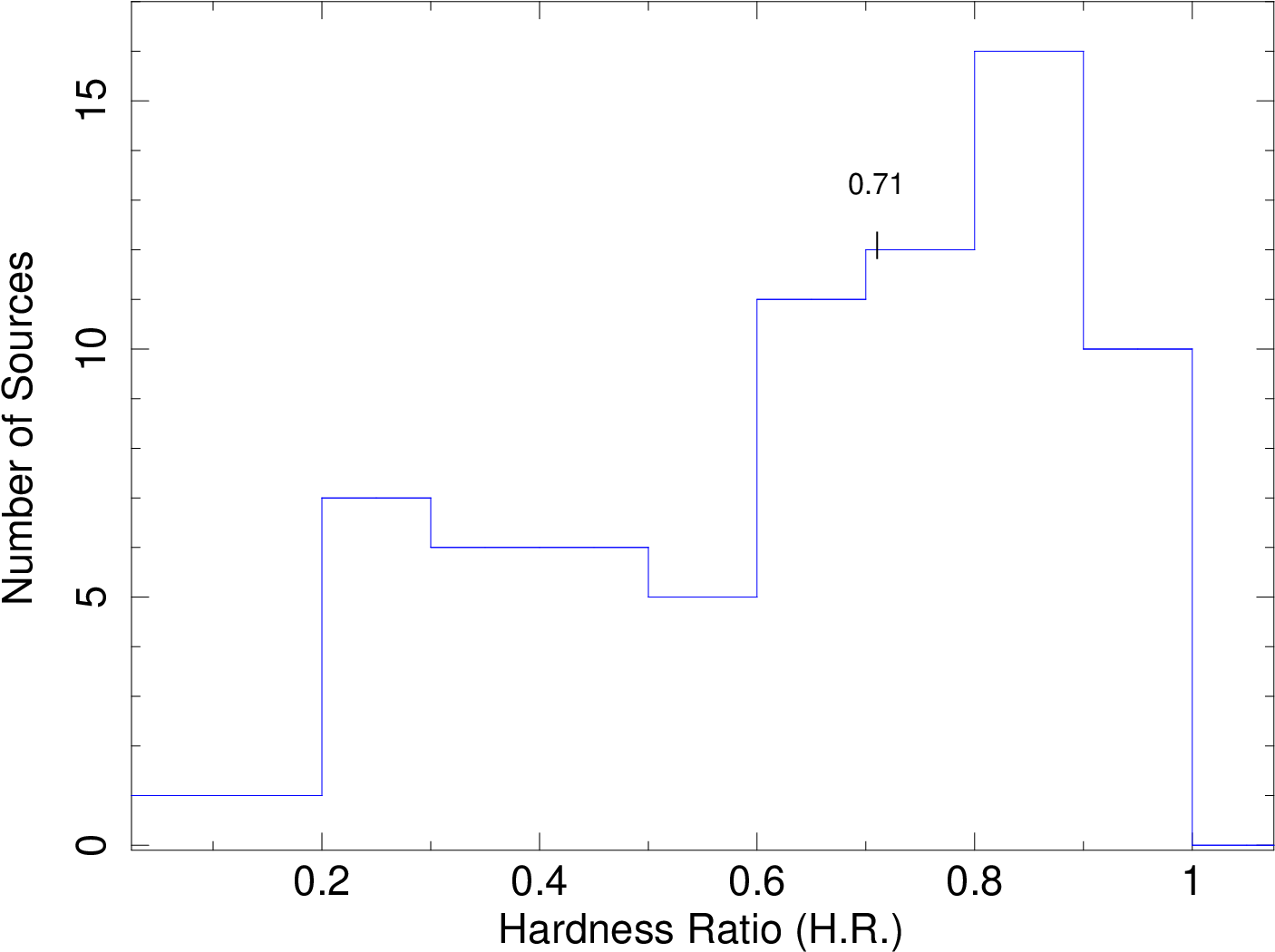} 
\caption{
Properties of Chandra Norma sources based on 0.5-7 keV events (Table 1).
Top: 
Background-subtracted median event energies.
The median is E$_{50}$ = 2.88 keV.
Bottom:
Background-subtracted Hardness Ratio, H.R. = counts(2-7 keV)/counts(0.5-7 keV).
The median is H.R. = 0.71.
}
\end{figure}

\subsection{IR and Optical Counterparts}

Candidate optical and IR counterparts whose positions are offset 
by $\leq$3$''$ from the X-ray positions are noted in Tables 1 and 2.
A total of 32 Chandra sources have potential 2MASS near-infrared (nIR) matches,
but matches for CXO 2,9,10,74 are uncertain. 
The candidate 2MASS matches for CXO 2 and 9 have offsets 
$>$3$''$ and the 2MASS images for CXO 9 show a marginally
resolved pair. The 2MASS fields near CXO 10 and 74 
reveal multiple nIR sources near the X-ray position.
Table 2 lists 49 candidate 2MASS counterparts for the XMM detections,
of which XMM 1,12,86,89  are uncertain.
The 2MASS position offsets for XMM 12 and 86 are $>$3$''$                                                                        \
and the 2MASS fields near XMM 1 and 89 show multiple sources
near their X-ray positions.
A search of the Vista Variables in the Via Lactea (VVV) DR2
catalog (Minniti et al. 2010) using the Chandra and XMM-Newton
positions found potential nIR matches for many of the X-ray
sources and those of particular interest for purposes of
YSO identification are identified below (Sec. 3.4). 
A search of the WISE All-Sky Survey source catalog returned
20 mid-IR sources within 3$''$ of the Chandra positions.
A similar search using the XMM-Newton positions found 
22 candidate WISE matches of which 14 were also detected
by Chandra.

The Gaia DR3 archive returned at least one potential counterpart 
for 41 Chandra sources where the Gaia position lies inside the 
Chandra 3$\sigma$ position error ellipse (Table 1).
None is is flagged by Gaia as a QSO or galaxy candidate.
Particularly useful are the Gaia parallax distances for 
the emission line stars Sz 136 (861.5$^{+18.3}_{-17.5}$ pc) and 
Sz 137 (879.5$^{+49.5}_{-44.5}$ pc). These reddened YSOs provide an
accurate benchmark for the cloud distance.
Also, 57 candidate Gaia DR3 counterparts within 3$''$ of
XMM-Newton source positions were found (Table 2). 
 
In addition, 19 Chandra sources have Hubble Space Telescope
v. 2.3.2 Guide Star Catalog (HST GSC) counterparts, 
of which 7 have class=0 (star) classifications and
12 are classified as class=3 (non-star). However,
these classifications should be treated with caution
since nebulous YSOs can be misclassified as non-stellar (e.g. Re 13).

\subsection{Sources with Near-Infrared Excesses}

Figure 3 shows color-color (H$-$K$_{s}$,J$-$H) and
color-magnitude  (J$-$H,K$_{s})$ diagrams for 
X-ray detections having 2MASS counterparts detected
at J,H, and K$_{s}$ with good-quality photometry.
Of the 26 Chandra sources with good
photometry, four have colors with nIR exesses compared
to unreddened classical TTS (Meyer et al. 1997).
These include Sz 136 (CXO 64) and Sz 137 (CXO 73), 
strengthening their status as PMS stars. 
The 2MASS counterpart for CXO 47 has a modest nIR excess 
and its projected position lies within the darkest part of
the Sa 187 filament, suggesting that it may be a PMS star.
Its HST GSC counterpart is classified as class=3 (non-star)
but it is probably a distant star since there is a faint Gaia DR3
source ($g$ = 18.37 mag) offset 0$''$.73 from the CXO position. 
The Gaia DR3 parallax distance d = 1.37 kpc and the median 
geometric distance from Bailer-Jones et al (2021)  
d = 1.95 (1.34 - 3.80) kpc place it behind the cloud.
CXO 67 has a candidate 2MASS counterpart offset 1$''$.9 from 
the X-ray position with a nIR excess 
but its H-band photometry has a confusion flag set
so its colors may not be reliable.

Figure 3 also includes 41 XMM-Newton sources
with good quality 2MASS photometry but reveals only one
additional nIR excess source that was not detected by Chandra.
This source XMM J163332.91$-$4459045 fell outside the ACIS-I FoV and
its counterpart 2MASS J16333290$-$4459052 
shows a small nIR excess so it is of potential interest as 
a YSO candidate and is thus included in Figure 3.
An additional five VVV nIR excess sources were found
for X-ray sources without a 2MASS counterpart as identified in
Tables 1 and 2. The VVV counterpart of CXO 34 is ambiguous because 
of multiple VVV sources near the X-ray position. 
Also, CXO 60 and 71 have little if any excess compared
to normally reddened stars (Fig. 3).

\begin{figure}
\figurenum{3}
\includegraphics*[width=7.5cm,angle=-90]{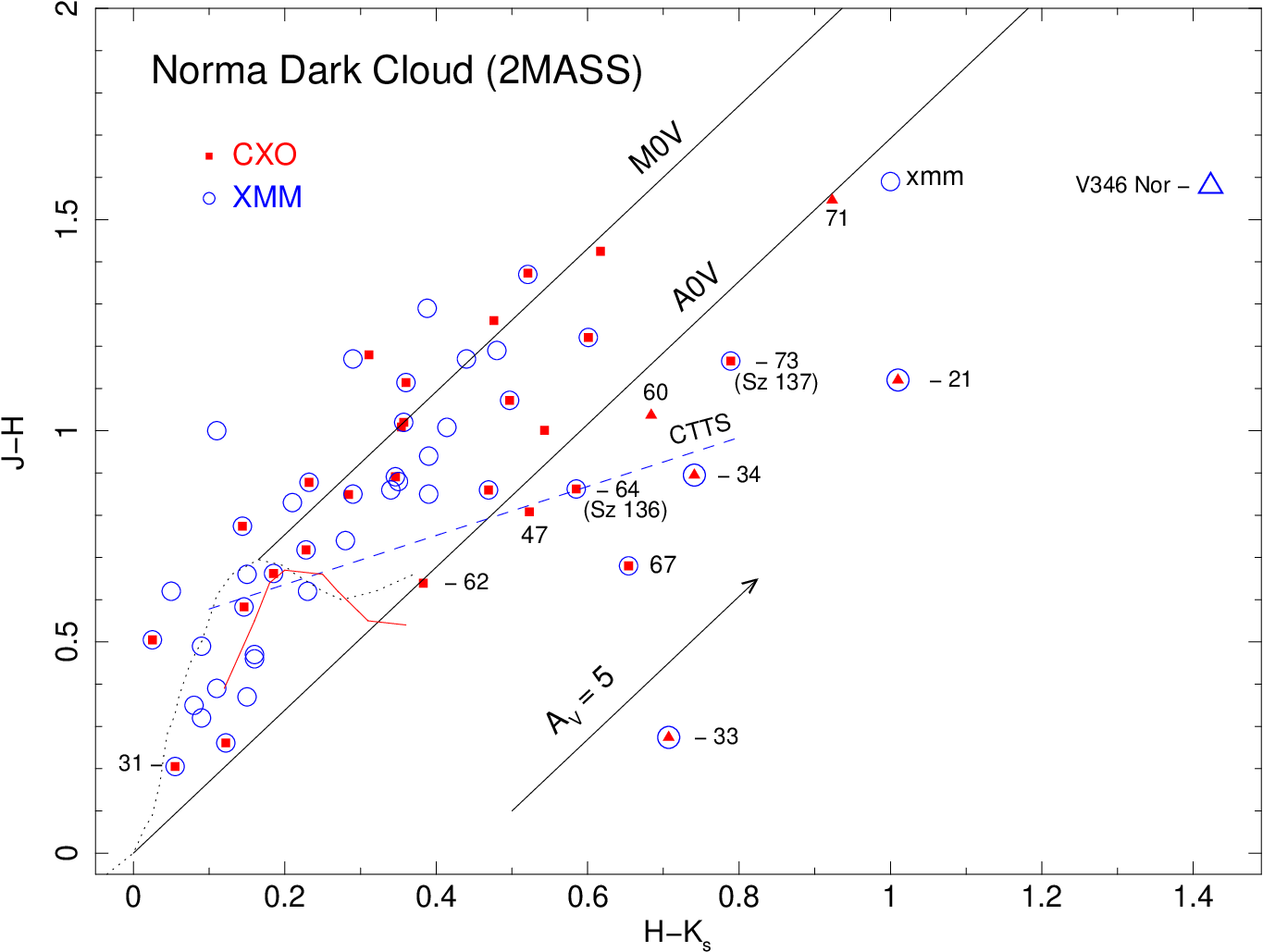} \\
\includegraphics*[width=7.5cm,angle=-90]{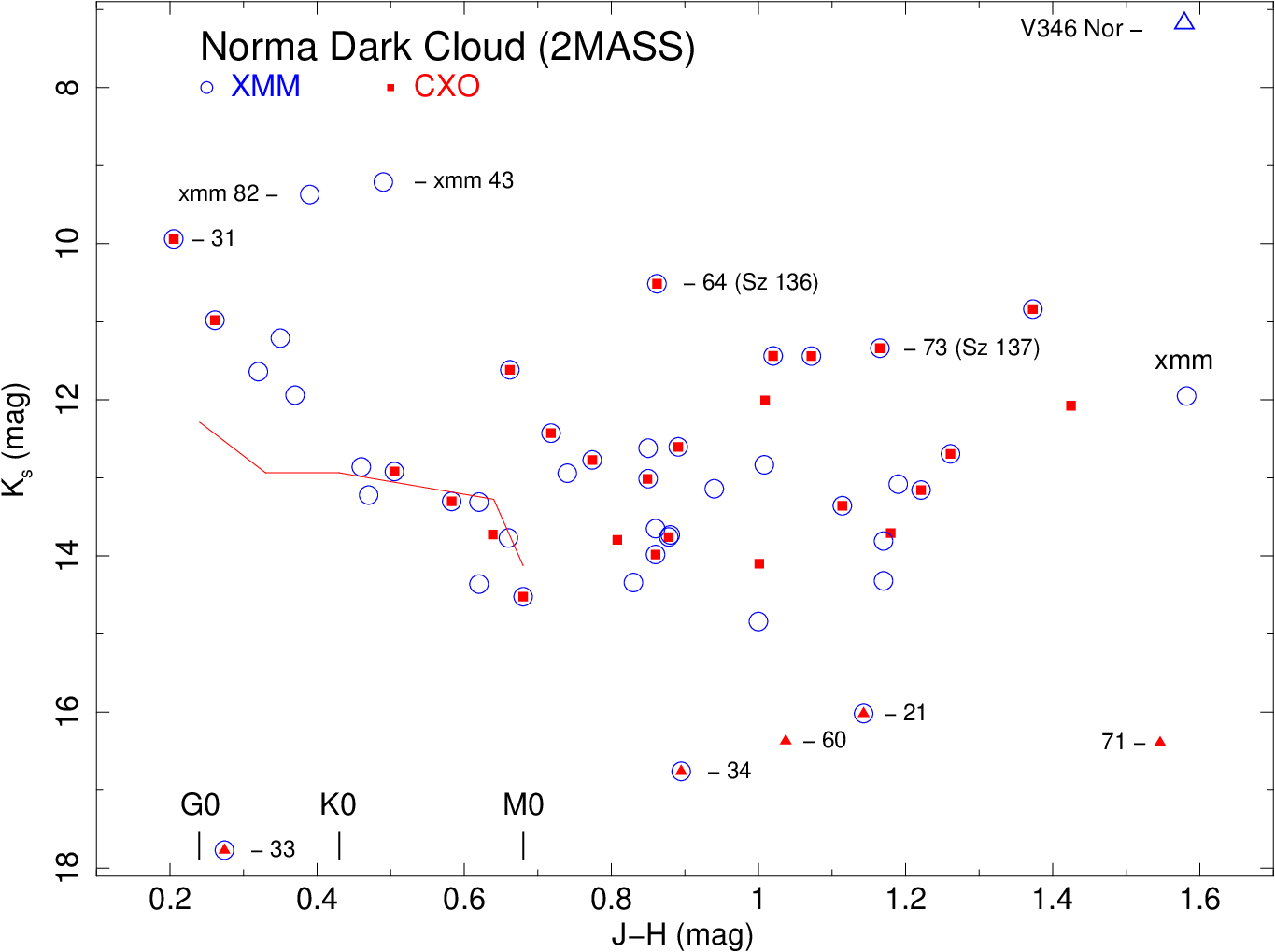}
\caption{
Top: 2MASS color-color diagram for all Chandra sources detected in the 
Norma dark cloud and XMM-Newton sources captured in the Chandra ACIS-I
field-of-view. Only 2MASS sources detected in all three bands and lying within
3$''$ of the X-ray positions are shown. 
Red triangles mark sources with nIR excesses from the VVV Survey 
(Minniti et al. 2010) that are not listed in the 2MASS catalog.
The VVV colors and magnitudes are based on the default 2$''$ aperture (Ap3).
The VVV match for CXO 34 is ambiguous due to multiple VVV sources near the 
X-ray position. 
Numbers correspond to CXO sources in Table 1. 
CXO 31 = HD 328329 (A2 star). CXO 71 is questionable for a nIR excess.
The point labeled xmm corresponds to
source XMM J163332.91$-$445904.5 which lies outside the Chandra 
FoV and has a 2MASS counterpart J16333290$-$4459052 with a marginal nIR excess.     
For comparison, the FUor-like star V346 Nor is shown but was undetected in X-rays.
The solid black lines show colors for normally-reddened M0V and A0V stars 
(Bessell \& Brett 1988). The curved dotted black line at lower-left is the 
unreddened main-sequence. The blue dashed line corresponds to 
intrinsic (unreddened) colors of CTTS from Meyer et al. (1997).
The solid red line corresponds to intrinsic colors of 5-30 Myr old 
G8-M6 stars from Pecault \& Mamajek (2013).
Bottom: Same as above, only J-H color versus K$_{s}$. 
The red line shows intrinsic values for G0 - M0 stars at 
700 pc (Pecault \& Mamajek 2013).
The bright K$_{s}$ sources XMM 43 and 82 are foreground stars. \\
}
\end{figure}

\subsection{On-Cloud and In-Cloud Sources}

The POSS2 IR image in Figure 1 shows 25 Chandra and 20 XMM-Newton sources 
that are projected onto the darkest part of the Sa 187 filament or its margin.
They are identified as on-cloud (o) in the first column of Tables 1 and 2.
Nine of the twenty XMM-Newton on-cloud
sources were also detected by Chandra (CXO 36,45,50,54,58,61,64,73,74)
giving 36 unique on-cloud sources.
The median value of
E$_{50}$ of the on-cloud Chandra sources is
2.67 keV and the median hardness ratio is 0.68.
These values are nearly the same as for the entire Chandra sample
(E$_{50,med}$ = 2.88 keV and H.R.$_{med}$ = 0.71).
Eleven of the on-cloud Chandra sources have potential 2MASS counterparts
with a range of K$_{s}$ = 10.51 (CXO 64 = Sz 136) to
K$_{s}$ = 14.46 (CXO 74).

We have also used geometric parallax distances from the 
Gaia DR3 archive\footnote{https://gea.esac.esa.int/archive} 
to identify X-ray sources that are likely in the cloud, 
and foreground sources. Those sources with Gaia DR3
distances in the range 500 - 1000 pc are considered as 
in-cloud (i) and those with d $<$ 500 pc are classified 
as foreground (f) as identified in Tables 1 and 2.
Using the above criteria and accounting for CXO and XMM-Newton
duplicate detections we find 7 foreground sources and
27 in-cloud sources. The nearest foreground object is
CXO 1 = XMM 11 at a Gaia distance of 121 pc.
These identifications are no doubt incomplete since not
all of the X-ray sources have Gaia distances.

\section{Comments on Specific Objects}

\subsection{V346 Nor}

The nebulous star V346 Nor erupted optically sometime between 1978-1983
(Graham \& Frogel 1985). Evidence that it is a young star includes an 
IR excess, CO outflow, and the associated jet-like outflow HH 57 (RN08).
Its IR spectral energy distribution (SED) resembles a 
class I protostar (Prusti et al. 1993) and it has historically
been classified as an  FU Orionis star, or FUor (Graham \& Frogel 1985;
Sandell \& Weintraub 2001; RN08). 
A more recent study by Connelley \& Reipurth (2018) noted that V346 Nor 
has some spectral characteristics that are not present in classical FUors and 
reclassified it as a ``peculiar'' star, albeit with some FUor characteristics.

V346 Nor was captured on-axis in the Chandra observations but was not 
detected (Fig. 4-top) and was also undetected by XMM-Newton. 
We use the lower background Chandra ACIS-I data to estimate an upper limit on 
its intrinsic X-ray luminosity L$_{x}$. We assume a 4 count detection threshold in 
the merged observations (0.067 c ks$^{-1}$), and a thermal plasma spectrum with
kT $\approx$ 3 keV. Hot plasma at similar or higher temperatures is usually
present in X-ray detected FUors including the prototype FU Orionis 
(Skinner et al. 2006; 2010). Cooler plasma from accretion shocks may also be 
present but is masked by high absorption.
The extinction of V346 Nor was found to be variable by 
K\'{o}sp\'{a}l et al. (2020). Based on the optical depth of 
the Si 9.7 $\mu$m absorption feature they obtained A$_{\rm V}$ = 16.4 mag
from a 2016 observation but this Si feature in low-resolution Spitzer 
spectra acquired in 2004 and 2006 gives  A$_{\rm V}$ = 7.3-8.8 mag.
Adopting  A$_{\rm V}$ = 16.4 mag yields an unabsorbed flux limit
F$_{x,unabs}$(0.5-7 keV) = 2.8$\times$10$^{-15}$ ergs cm$^{-2}$ s$^{-1}$,
and L$_{x}$ = 3.35$\times$10$^{23}$d$_{pc}^{2}$ ergs s$^{-1}$.
If it lies at the nominal cloud distance of 700 pc then
log L$_{x}$ = 29.21 ergs s$^{-1}$. 
If the average cloud extinction  A$_{\rm V}$ = 7.6 mag is
assumed then the above limit decreases to 
log L$_{x}$ = 29.04 ergs s$^{-1}$ (d=700 pc). 
The above L$_{x}$ limits are at the low end of the observed range 
for FUors and FUor-like stars,
but some have gone undetected at this level. For example, the FUor 
V733 Cep was undetected by Chandra at log L$_{x}$(0.3-8 keV) $\leq$ 28.95 ergs s$^{-1}$
(Skinner \& G\"{u}del 2020a). In contrast, some FUors are quite luminous in 
X-rays including the prototype FU Ori at log L$_{x}$ $\approx$ 31 ergs s$^{-1}$
(Skinner et al. 2010).

\subsection{Re 13 and HH 56}

The obscured star Re 13 illuminates a reflection nebula 
and has a rising SED characteristic of 
a class I protostar (Prusti et al. 1993). 
However, Alvarez et al. (1986) reported H$\alpha$ emission
in its reflection spectrum so the hidden object may be
an emergent cTTS in the class I/II transition stage.
There is no Gaia DR3 counterpart to Re 13.
There is no significant Chandra X-ray emission at the positions of 
Re 13 or MMS 1 (Fig. 4-top right), nor is any detected by XMM-Newton. 

Re 13 is thought to drive HH 56 located $\approx$40$''$ away
(Reipurth et al. 1997). We find no X-ray emission at the HH 56 
bow shock position (RN08) in the Chandra (Fig. 4-top) or XMM-Newton images. 
Typical HH X-ray luminosities are
L$_{x,{\rm HH}}$ $\approx$ 10$^{-4}$ L$_{\odot}$,
or log  L$_{x,{\rm HH}}$ $\approx$ 29.6 ergs s$^{-1}$ (Raga et al. 2002).
The maximum postshock temperature for a strong HH bowshock
formed at typical HH jet velocities of a few 
hundred km s$^{-1}$ is kT$_{s}$ $\approx$ 0.1 - 0.3  keV
(Raga et al. 2002; Pravdo et al. 2001; Favata et al. 2002). 
The Chandra detection limit at such low
temperatures depends sensitively on the intervening
absorption. Even a moderate extinction
of  A$_{\rm V}$ $>$ 3 mag toward HH 56 would be
sufficient to render it undetectable by Chandra
assuming kT$_{s}$ = 0.2 - 0.3 keV and d $\approx$ 700 pc.
This conclusion also applies to HH 57 near V346 Nor
which was also undetected (Fig. 4-top right). 

\subsection{Millimeter Sources MMS 1-5 and Environs} 

The compact millimeter sources MMS 1-5 listed in NC05
were not detected (Fig. 4-top) and MMS 6 fell 
outside the X-ray FoV.
As noted above, MMS 1 (Re 13) may be a class I
protostar and MMS 4 as well (NC05; Moreira et al. 2000).
Class I protostars are in some cases detected in 
X-rays only during large flares and are otherwise
X-ray quiet (e.g. Imanishi, Koyama, \& Tsuboi 2001). 
Thus, persistent X-rat time monitoring may be needed 
to detect them.

We note the source XMM 81 located SE of 
MMS 5 (Fig 4-top left). 
It is a rather bright XMM-Newton source 
but not a significant Chandra detection, 
although a few ACIS-I counts are visible.  
This suggests X-ray variability.
It has a 2MASS match with H-K$_{s}$ = 2.01 
and J $\geq$ 18.1 mag but no Gaia DR3 counterpart.
XMM 81 lies near a EPIC pn CCD gap so we extracted
MOS1 and MOS2 light curves and spectra.  The light
curves (0.5-7 keV) show no flares and when rebinned 
to 3 ks bins there are no fluctuations greater than
2$\sigma$. A simultaneous 1T apec fit of the 
MOS spectra gives 
N$_{\rm H}$ = 1.28 (1.00 - 1.57) $\times$ 10$^{22}$  cm$^{-2}$
(A$_{\rm V}$ $\approx$ 6.7 mag),  kT = 3.01 [2.43 - 4.05] keV, and
F$_{x,abs}$(0.5-7 keV) = 6.01 $\times$ 10$^{-14}$ ergs cm$^{-2}$ s$^{-1}$.
On the basis of its X-ray variability, high plasma 
temperature, signficant absorption, and projected position near MMS 5, 
XMM 81 may be a YSO.

\begin{figure}
\figurenum{4}
\includegraphics*[width=8.0cm,height=8.0cm,angle=0]{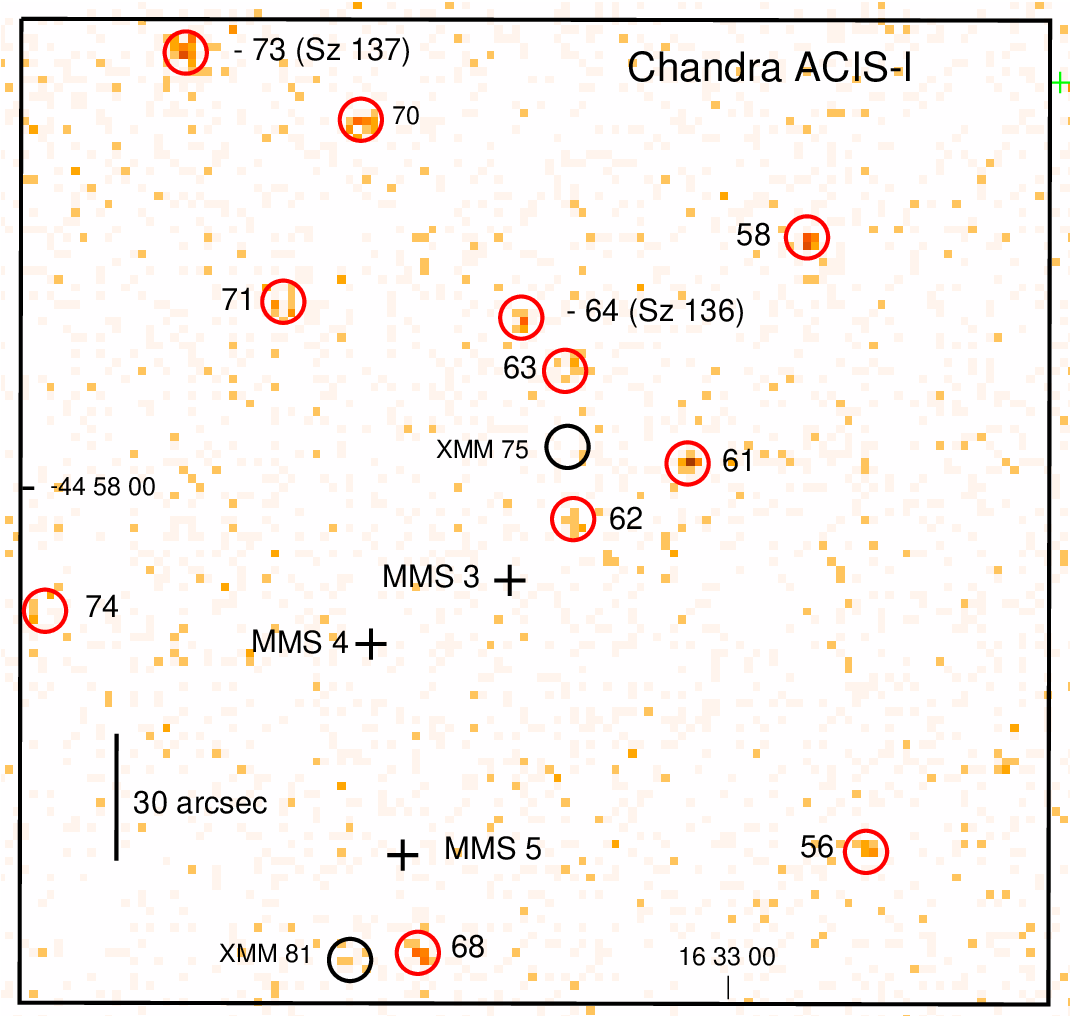}
\hspace{0.3cm}
\includegraphics*[width=8.0cm,height=8.0cm,angle=0]{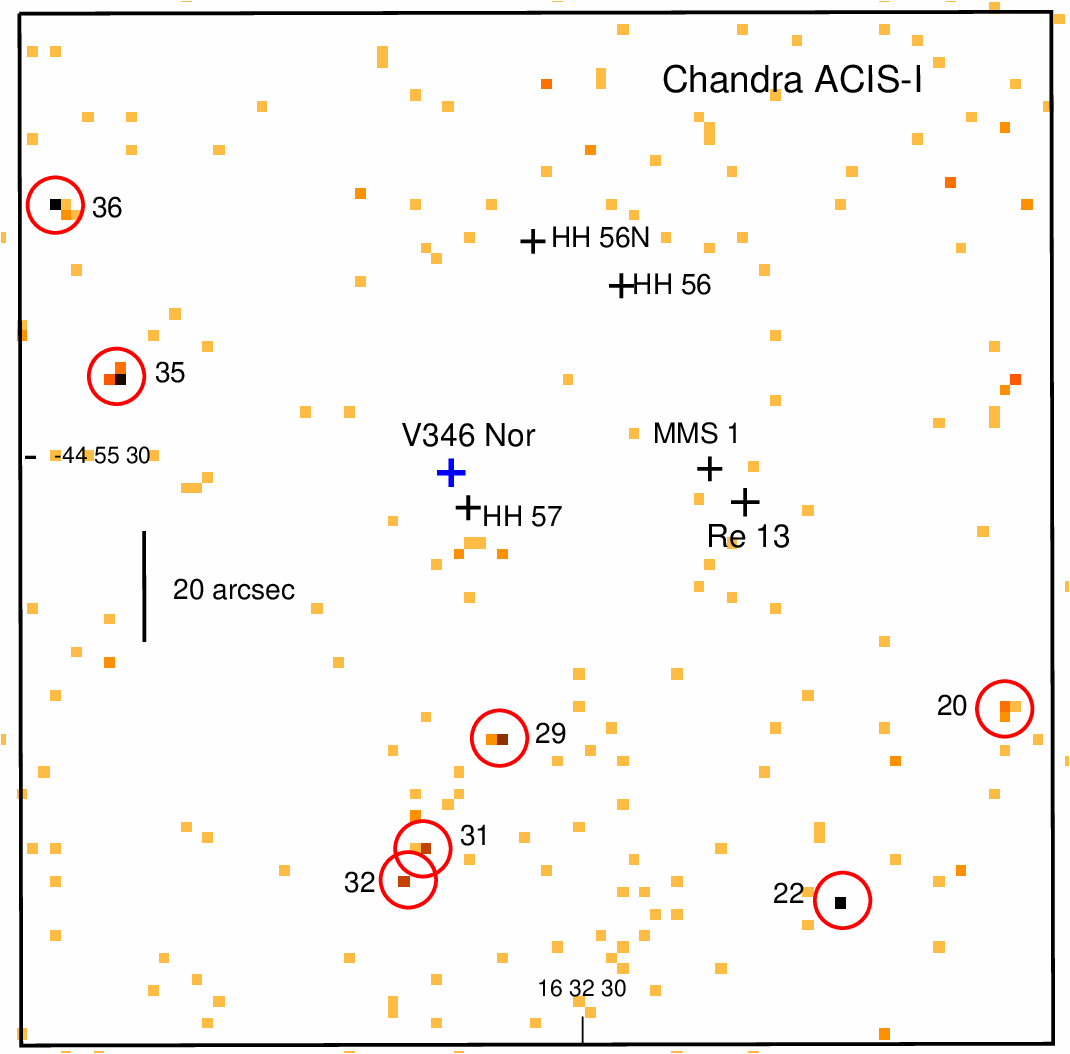} \\
\includegraphics*[height=8.0cm,angle=-90]{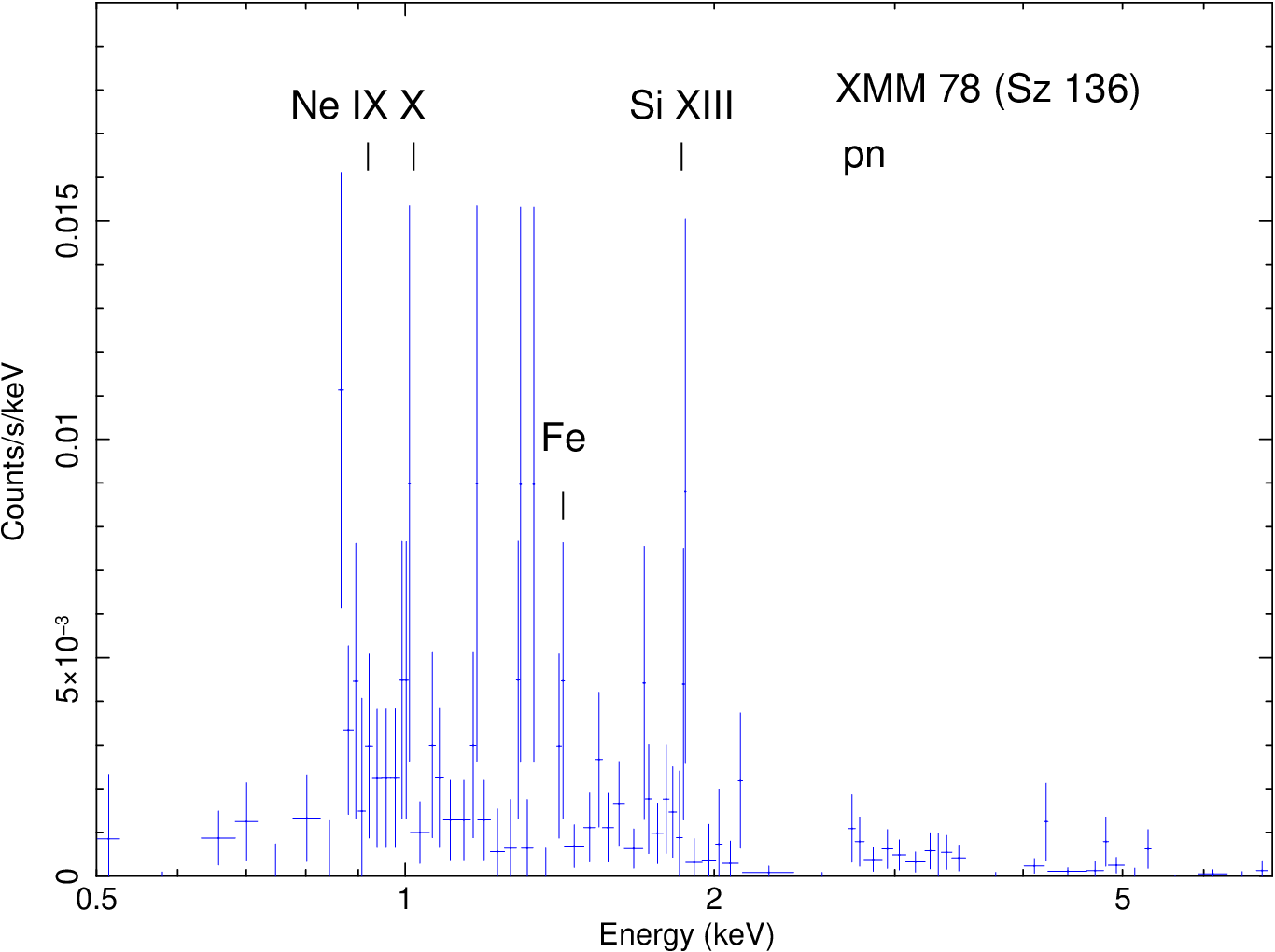}
\hspace{0.3cm}
\includegraphics*[height=8.0cm,angle=-90]{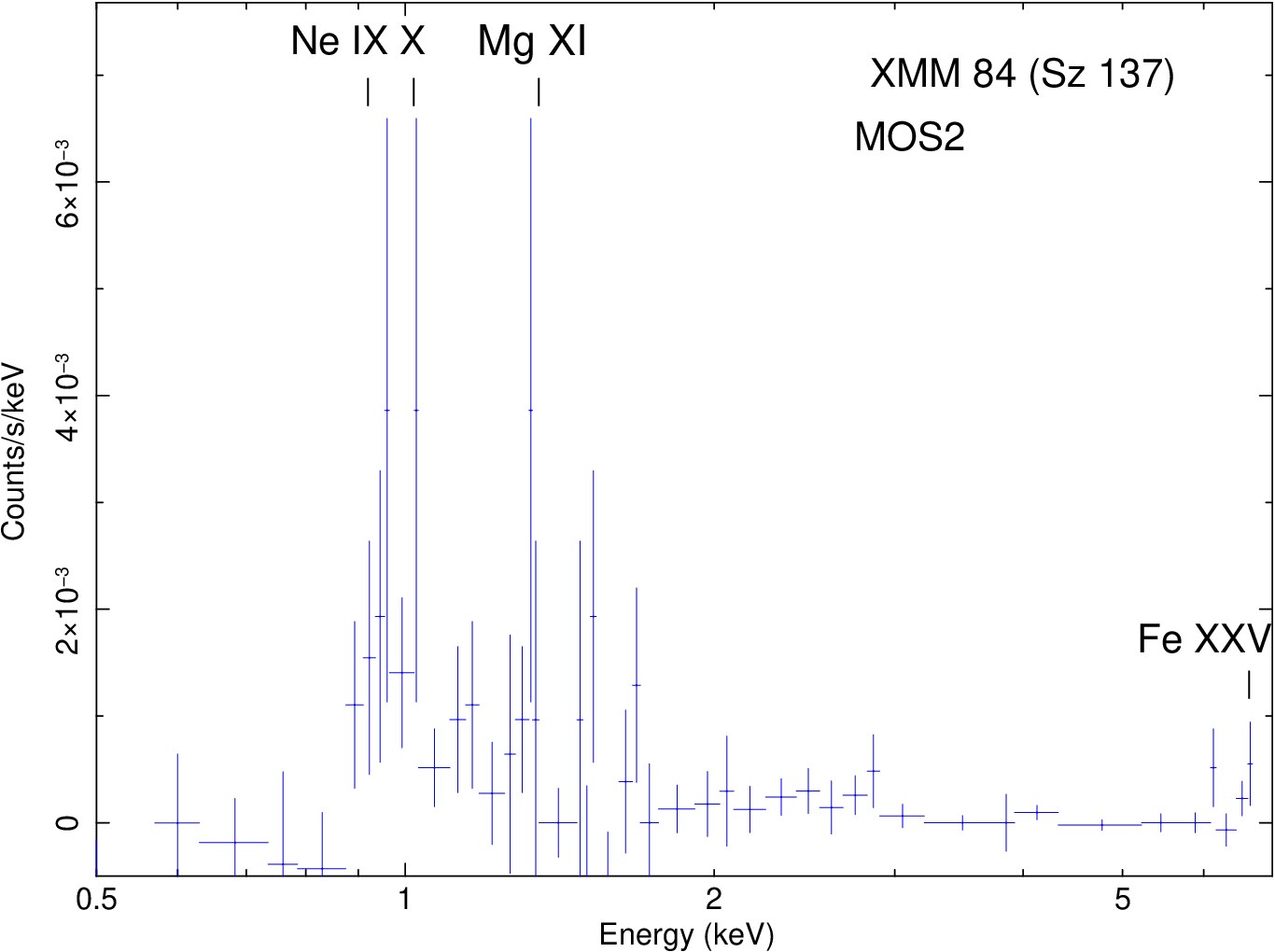}
\caption{
Top Left:~Binned Chandra ACIS-I image (0.5-7 keV) near Sz 136 showing Chandra (red)
sources, all of which were detected by XMM-Newton except CXO 62 and CXO 74.
The XMM-Newton (black) X-ray sources XMM 75 and XMM 81  were not
detected by Chandra, but very faint Chandra emission is visible at XMM 81.
Crosses mark positions of undetected objects from RN08.
Top Right: Same as at left except showing the region near V346 Nor.
Sources CXO 31 and CXO 36 were also detected by XMM-Newton. \\
Bottom: Lightly binned ($\geq$2 cts/bin) XMM-Newton spectra with possible 
emission lines marked. \\
Left: EPIC pn spectrum of Sz 136. 
Right: EPIC MOS2 spectrum of Sz 137 showing faint Fe XXV line.
}
\end{figure}

\subsection{Sz 136} 
The H$\alpha$ emission line star Sz 136 = CXO 64 
(Fig. 4-top left) was detected in only one of the four 
Chandra observations (ObsId  23387) at a net rate of
 1.04$\pm$0.33 c ks$^{-1}$. Since the exposure times of the four
observations were nearly the same (Sec. 2.1) Sz 136 is considered variable. 
Inspection of the Sz 136 low-count ACIS-I light curve for ObsId 23387 
shows no significant variability or flares.

The XMM-Newton detection of Sz 136 (XMM 78) provides more counts
than the faint Chandra detection so its EPIC pn spectrum was extracted.
An extraction region of radius 10$''$ was used to minimize 
contamination from a nearby source lying 
$\approx$16$''$ to the south and visible as CXO 63 in Fig. 4-top. 
A lightly binned EPIC pn spectrum (Fig. 4-bottom left) reveals weak line-like 
features near 0.9 keV (Fe or Ne) and 1.86 keV (Si XIII) indicative
of optically thin thermal plasma emission. 

A two-temperature optically thin thermal plasma model
(2T apec) provides a better fit of the EPIC pn spectrum than an
isothermal 1T apec model (Table 3) but the 2T model does not
tightly constrain the temperature of the hot plasma component (kT$_{2}$).
The 2T apec fit column density N$_{\rm H}$ corresponds to
A$_{\rm V}$ = 4.1 [1.7 - 5.7; 1$\sigma$ range] mag and
the unabsorbed flux gives a luminosity
log L$_{x}$(0.5-7 keV) = 30.64 ergs s$^{-1}$ (d = 861.5 pc).
For comparison with XMM-Newton, a 1T apec fit of the
low-count Chandra spectrum of Sz 136 (ObsId 23387)
with a 1T apec model and freezing N$_{\rm H}$ at the value
obtained from the 1T apec EPIC pn fit for consistency 
gives kT = 0.73 (0.40 - 1.18) keV,
F$_{x,abs}$(0.5-7 keV) = 1.02 $\times$ 10$^{-14}$ ergs cm$^{-2}$ s$^{-1}$,
and log L$_{x}$(0.5-7 keV) = 31.00 ergs s$^{-1}$.
The above L$_{x}$ values for Sz 136 place it at the high end of the
range observed for TTS based on a comparison with the values
determined for TTS in the Orion Nebula Cluster (ONC) by Preibisch et al. (2005).

\subsection {Sz 137 and Environs}

The emission line star Sz 137 = CXO 73 (Fig. 4 top-left) was
only detected in one Chandra observation (ObsId 26283)
at a net rate of 3.32$\pm$0.58 c ks$^{-1}$ averaged over the observation. 
Its ACIS-I light curve reveals a decline in count rate
from $\approx$9 c ks$^{-1}$ to $\approx$1 c ks$^{-1}$ (0.5-7 keV).
An exponential fit of the decay gives an e-folding time of $\approx$6 ks
so the star was evidently caught during the decay phase of a flare.

Sz 137 was also detected by XMM-Newton (XMM 84).
It fell in a  EPIC pn camera CCD gap so its MOS1 and MOS2 
spectra were extracted and fitted simultaneously. 
The lightly binned spectrum (Fig. 4 bottom-right) shows emission
features near 1 and 1.2 keV that are probably faint Ne lines.
Also, a faint Fe XXV line is visible near 6.64 keV, a signature of
very hot plasma with maximum line power at T$_{max}$ = 63 MK.
A 2T apec fit is slightly better than 1T apec but does not 
tightly constrain the hot component temperature (Table 3).
The extinction determined from N$_{\rm H}$ is
equivalent to A$_{\rm V}$ = 4.7 [2.3 - 6.4] mag.
At the Gaia DR3 distance of 879.5 pc the unabsorbed MOS flux
gives log L$_{x}$(0.5-7 keV) = 30.56 ergs s$^{-1}$, nearly
the same as for Sz 136.
A fit of the Sz 137 Chandra spectrum for this observation 
returns a high but uncertain plasma temperature because of the 
variability and an absorbed flux 
F$_{x,abs}$(0.5-7 keV) = 6.93 [5.95 - 8.31; 1$\sigma$ range] $\times$ 10$^{-14}$ ergs cm$^{-2}$ s$^{-1}$.
This flux is nearly 7 $\times$ higher than measured by XMM-Newton 
11 months earlier.

It is worth noting that 
the extinctions A$_{\rm V}$ $\approx$ 4-5 mag derived from fits of the
X-ray spectra of Sz 136 and Sz 137 are somewhat
less than the mean cloud extinction A$_{\rm V}$ = 7.6 mag (NC05).
This provides support for previous studies which have noted
spatial variation in extinction across the Norma 
cloud including relatively transparent cavities with visible background
stars (Reipurth et al. 1997; Alvarez et al. 1986, 
Waldhausen \& Marraco 1993).

\subsection{HD 328329}

This A2e star was faintly detected as a soft X-ray source by Chandra in 
the merged observations (CXO 31) and by XMM-Newton (XMM 39).
The pn spectrum is soft with low absorption and most emission lies 
below 1 keV. A 1T apec thermal plasma fit of the pn spectrum yields 
kT $<$ 1 keV and 
F$_{x,abs}$(0.5-7 keV) = 0.18 [0.08 - 0.21] $\times$ 10$^{-14}$ ergs cm$^{-2}$ s$^{-1}$.
This flux is similar to that derived from the Chandra data. The combination of
a soft spectrum, low absorption, and its Gaia distance of 538 pc are consistent 
with this star lying on the near side of the dark cloud.
 
If HD 328329 is a young Ae star in the cloud then it may be
related to the class of pre-main sequence Herbig Ae/Be stars.
The origin of their X-ray emission has been strongly debated 
and in some cases may originate in unseen late-type companions
(Stelzer et al. 2006). But a few Herbig Ae stars show soft X-ray
spectra (kT $\leq$ 1 keV) that suggest intrinsic emission from 
the Ae star and not a late-type companion. These soft-spectrum
Herbig Ae stars include HD 163296 (Swartz et al. 2005), 
AB Aur (Telleschi et al. 2007b), and HD 100546 (Skinner \& G\"{u}del 2020b). 
There is no universal consensus on the origin of the soft X-ray
emission and several mechanisms have been proposed. These include
accretion shocks, current sheets, and soft coronal emission as
discussed in more detail in the above references.

\subsection{Bright X-ray Sources}

The brightest X-ray source detected by XMM-Newton is XMM 11 (= CXO 1)
which has a  2MASS counterpart (K$_{s}$ = 11.62 mag). 
Its Gaia DR3 parallax distance of 120.7 pc places it in the foreground
and not in the dark cloud. It is classified as a star 
in the HST GSC and VVV catalogs. More interesting and mysterious are    
the bright Chandra sources CXO 4 (= XMM 16) and CXO 21 (= XMM 34) discussed below.
Their X-ray spectra are shown in Figure 5.

\subsubsection{CXO 4  (= XMM 16)}

This is the brightest Chandra source but extensive catalog searches
found only two possible faint counterparts. There is a Gaia DR3 object
with g=20.86 mag at an offset of 2$''$.2 SE from CXO 4 which has no parallax 
measurement. It is not listed in the Gaia galaxy or QSO 
candidates catalogs (Bailer-Jones et al. 2023). 
There is also a faint VVV nIR source (not in 2MASS) offset 2$''$.0 SW
of CXO 4 with VVV class=star and J = 18.5 mag but no H or K$_{s}$ data.
The Gaia and VVV sources fall inside the CXO 4 3$\sigma$ X-ray position
error ellipse
which has semi-axis dimensions of 5$''$.1 $\times$ 4$''$.2.
The Gaia source Gaia DR3 J163202.45$-$445349.2 (positional uncertainty 0$''$.049) and 
VVV source (VVV J163202.13$-$445350.1) differ in position by 3$''$.5 
so it is unlikely they are the same object and it is uncertain which one 
(if either) corresponds to the X-ray source.

CXO 4 is variable with a rate of 4.6 - 6.5 c ks$^{-1}$ in 
ObsIds 23387 and 26283
and a much higher rate of 16.8 c ks$^{-1}$ in ObsIds 24511 and 27464.
The ACIS-I light curves for the high-state observations
show a nearly steady count rate and no flares.
The Chandra ACIS-I spectrum during high state reveals additional
hard emission above $\approx$3 keV compared to the low-state (Fig. 5-top left).
Such variable X-ray emission with  
increases in the hard band during high-state is typical 
of magnetically-active young stars and protostars, 
but does not conclusively establish a YSO classification 
without a confidently identified counterpart.

The heavily-binned XMM-Newton spectrum of XMM 16 (= CXO 4)
and XMM 34 (= CXO 21)
show no obvious lines other than a possible Ne IX peak near 
0.9 keV (Fig. 5-top right). A hard tail above 3 keV is visible in
both sources that signals very hot thermal plasma or possibly a 
nonthermal power-law (PL) contribution. When the XMM-Newton spectra 
are viewed under light binning (Fig. 5-bottom) they are noisy but 
a few possible lines are visible. In the the XMM 16 spectrum these
are the S XVI doublet (2.62 keV, T$_{max}$ = 25 MK) and 
Ar XVII (3.14 keV T$_{max}$ = 20 MK). The Ar XVII peak is also
visible in the Chandra ACIS-I spectrum  (Fig. 5-top left).
When viewed under heavier binning ($>$20 cts/bin) these
weak line-like features  wash out and                                                                                                  the spectra are dominated by continuum (Fig. 5 top-right).

Table 3 summarizes spectral fits of CXO 4 = XMM 16.
A fit of the ACIS-I high-state
spectrum with a 1T apec model requires very hot plasma 
but only  a lower bound kT $>$5 keV is obtained. Fits of the 
CXO 4 low-state spectrum with the 1T apec model yield
nearly identical results as the XMM 16 spectrum of this source
obtained 11 months earlier, converging to kT $\approx$ 2.3 keV.
Since the heavily-binned XMM 16 pn spectrum reveals a hard tail
above 3 keV, Table 3 includes a fit using s hybrid
thermal$+$nonthermal model (1T apec$+$PL).
This hybrid  model does not converge to a stable value of kT
so it was held fixed at kT = 0.5 keV. The PL component dominates
the emission measure with very little contribution from the 
thermal (apec) component.
As judged by the $\chi^2$ fit statistic there is no clear
preference for the 1T apec model over the 1T apec$+$PL model.

The models in Table 3 for CXO 4 = XMM 16
give similar N$_{\rm H}$  values equivalent to 
A$_{\rm V}$ $\approx$ 6-8 mag, which likely places the source
in the cloud or behind it. The average of the unabsorbed fluxes
from the different fits give
L$_{x}$(0.5-7 keV) = 1.3$\times$10$^{31}$(d$_{pc}$/700)$^{2}$ (low state)
and 2.6$\times$10$^{31}$(d$_{pc}$/700)$^{2}$ ergs s$^{-1}$ (high state).
The inferred L$_{x}$ at the nominal cloud distance of $\sim$700 pc
is at the high end of the range observed for TTS (Preibisch et al. 2005).

\begin{figure}
\figurenum{5}
\includegraphics*[height=8.0cm,angle=-90]{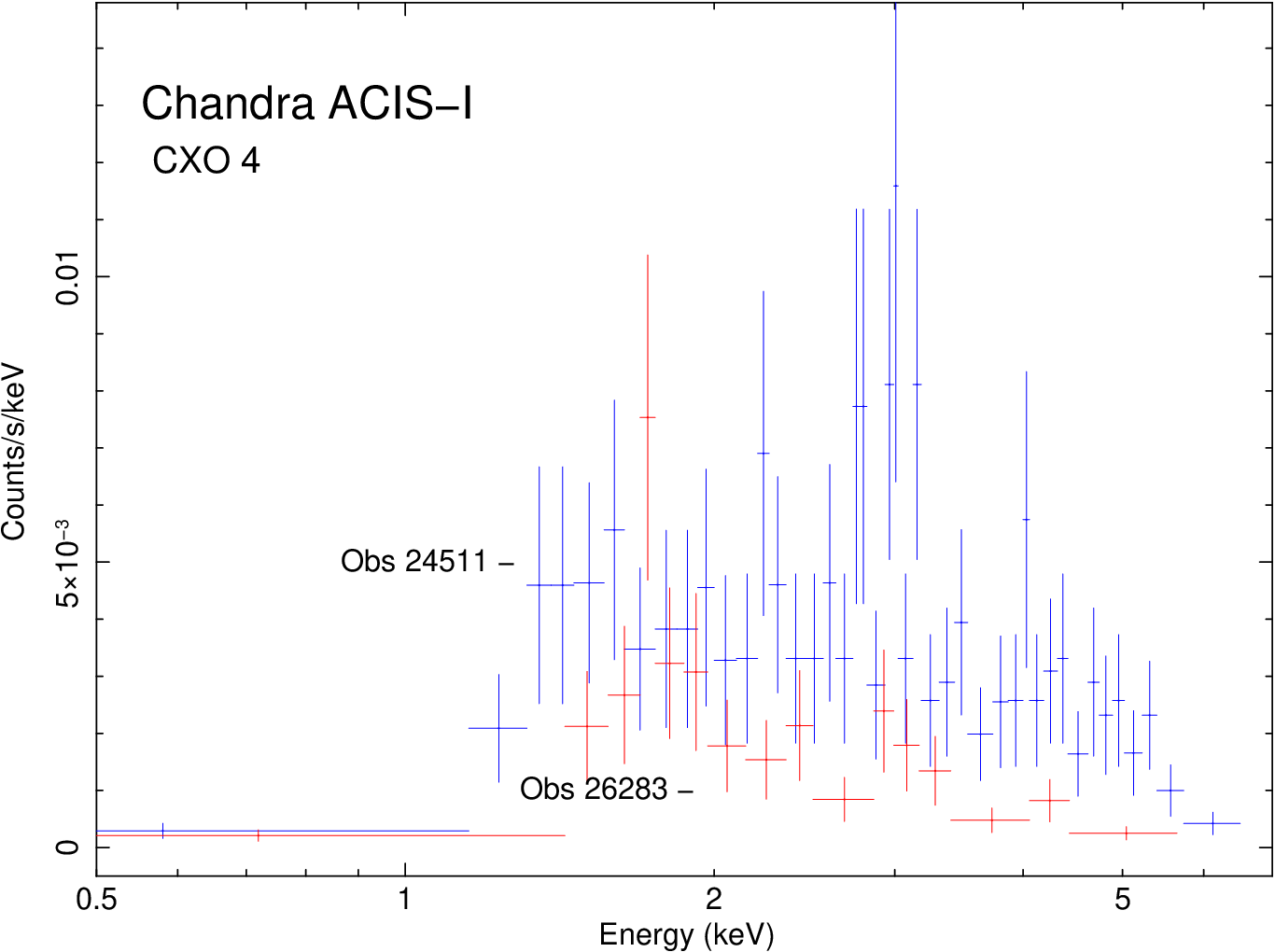}
\hspace{0.3cm}
\includegraphics*[height=8.0cm,angle=-90]{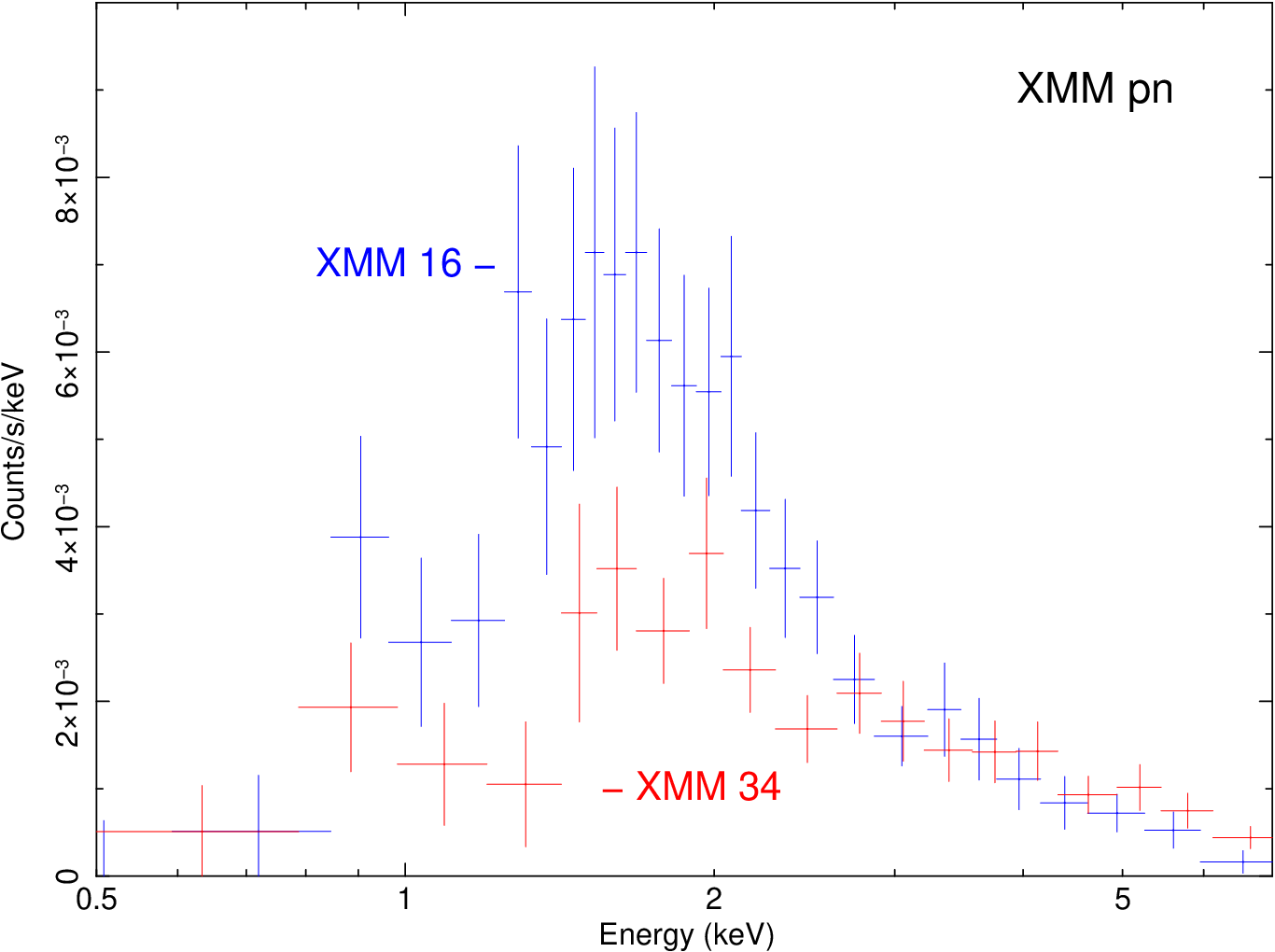} \\
\includegraphics*[height=8.0cm,angle=-90]{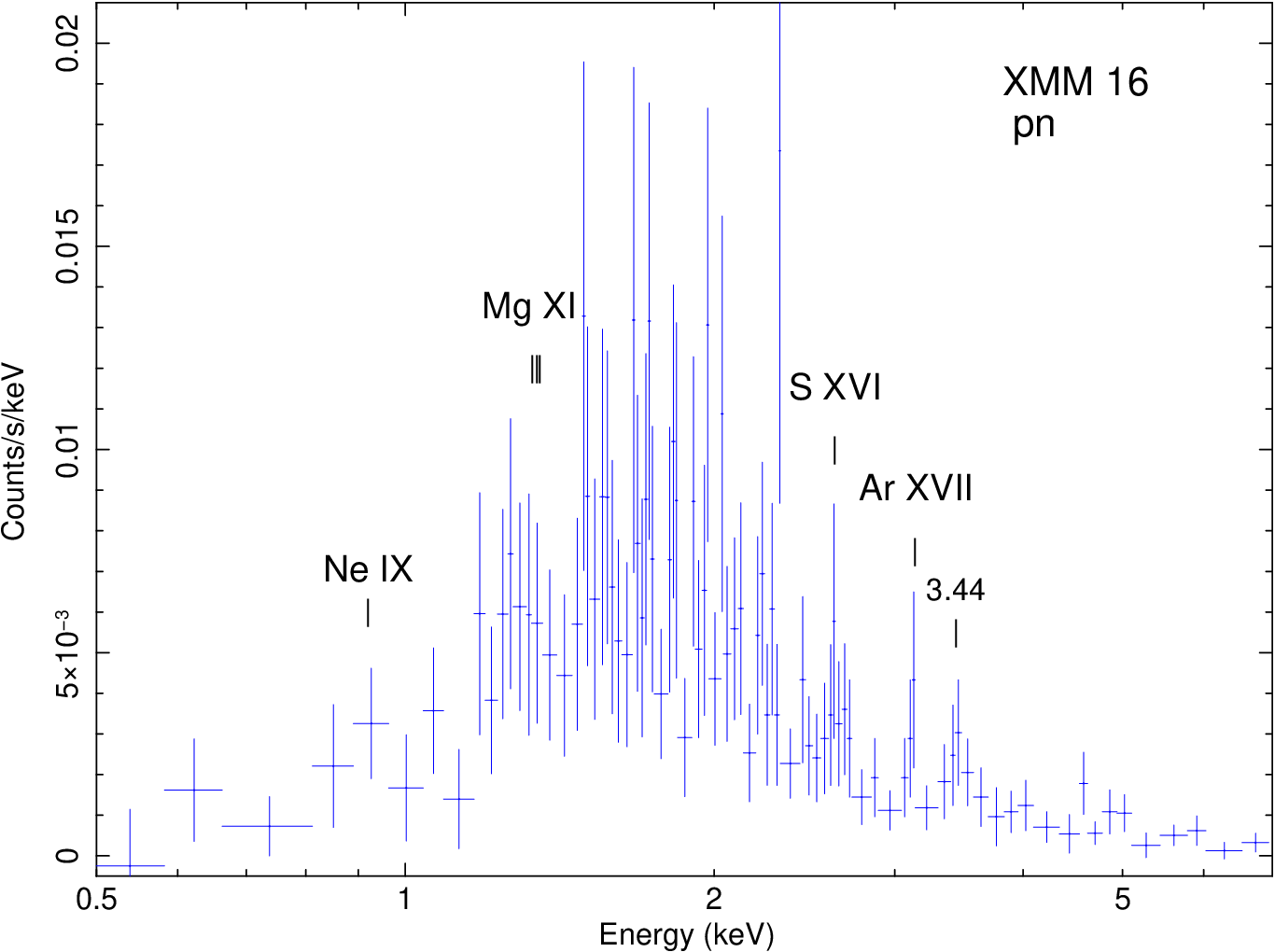}
\hspace{0.3cm}
\includegraphics*[height=8.0cm,angle=-90]{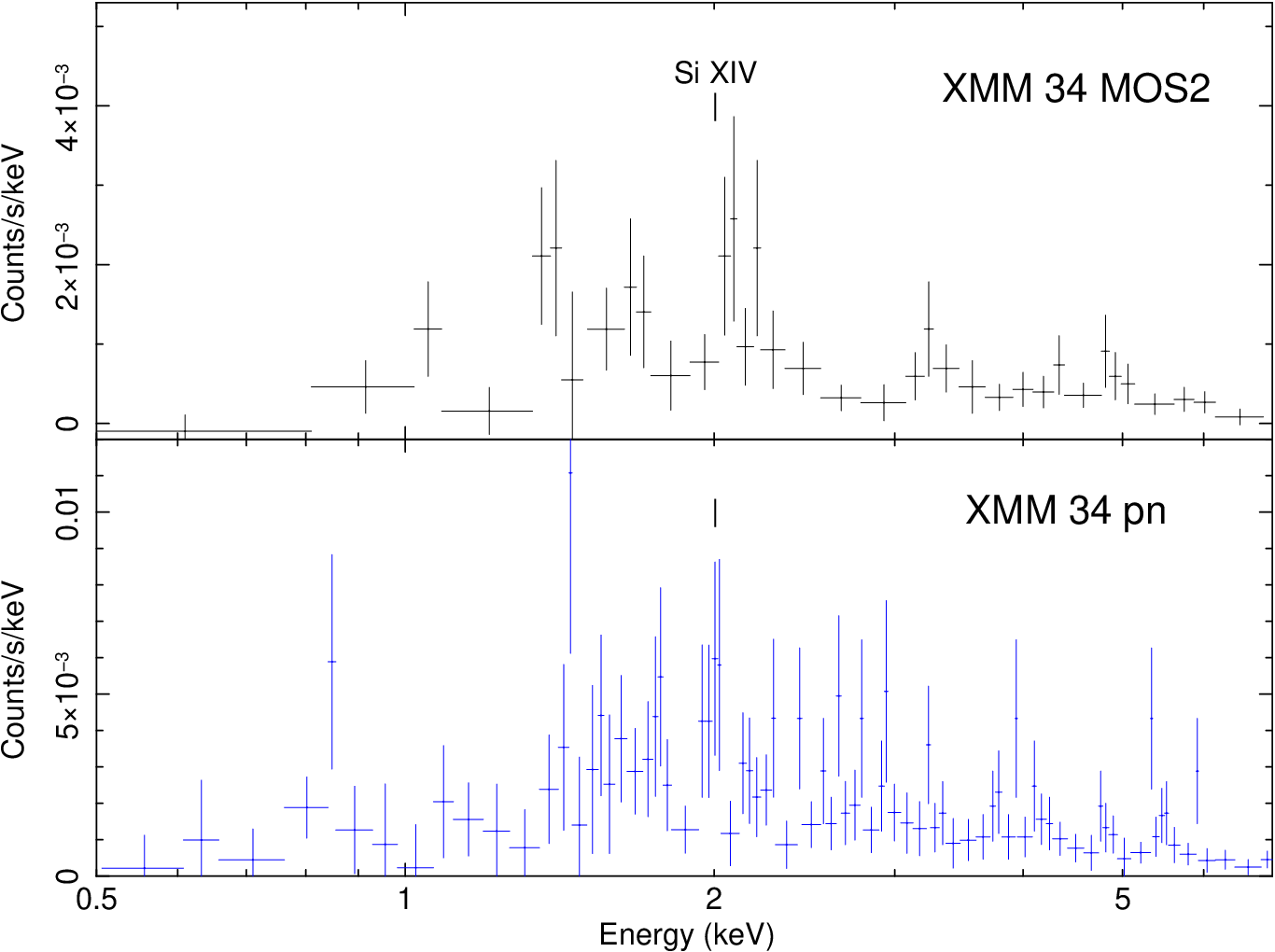}
\caption{
X-ray spectra with possible emission lines marked.
Top Left:~CXO 4 ACIS-I spectra in high state (Obs 24511, 227 net cts)
and low state (Obs 26283, 86 net cts) binned to $\geq$5 cts/bin. 
Top Right: XMM-Newton EPIC pn spectra of the bright sources XMM 16 (= CXO 4) 
and XMM 34 (= CXO 21) binned to $\geq$30 cts/bin.
Bottom Left: XMM-Newton EPIC pn spectrum of XMM 16 (= CXO 4), lightly binned to
$\geq$2 cts/bin to bring out possible lines. 
The feature at 3.44 keV is unidentified.
XMM 16 fell in a MOS CCD gap and no MOS spectra were extracted. \\
Bottom Right: XMM-Newton EPIC MOS2 and pn spectra of XMM 34 (= CXO 21), lightly binned
to $\geq$2 cts/bin.
}
\end{figure}

\newpage
\subsubsection{CXO 21 (= XMM 34)}

No optical match was found for CXO 21 but there is a  
nIR source in the VVV catalog offset by 0$''$.23 with
J,H,K$_{s}$ = 17.958, 16.815,, 15.793 mag (Ap3 photometry).
It has VVV class=galaxy but a significant nIR excess 
(Fig. 3) suggests that it may instead be an optically-obscured YSO.
It is visible in 2MASS and WISE All-Sky Survey images
but is not listed in their source catalogs.
However, it is listed in the unWISE catalog (Schlafly et al. 2019)
which incorporates W1 (3.4 $\mu$m) and W2 (4.6 $\mu$m) photometry
from the extended NEOWISE mission.
The unWISE flux densities convert to
W1 = 15.20, W2 = 14.04, and W1$-$W2 = 1.16 mag.
A fit of the IR SED using VVV K$_{s}$ and unWISE W1, W2  photometry
gives a slope $\alpha$ = $-$0.5$\pm$0.1 where 
$\alpha$ = d log($\lambda$F$_{\lambda}$)/d (log$\lambda$)
which is consistent with a class II YSO (cTTS) 
based on the classification system described by
Haisch et al (2001), Wilking et al. (2001), and references therein.
A second VVV source is present at a slightly larger offset 
of 1$''$.49 with VVV class=star which is $\approx$1 mag fainter at
J,H (no K$_{s}$ data). Its larger offset and faintness
make it a less likely counterpart to the X-ray source.

The lightly binned XMM-Newton spectra of the bright source
XMM 34 (= CXO 21) in Figure 5-bottom right show very few 
identifiable lines except for the 
Si XIV doublet (2.006 keV, T$_{max}$ = 16 MK).
Table 3 summarizes fits of this source using the same models
as discussed above  with similar results.
The 1T apec model requires a high but uncertain kT.
The various models yield
N$_{\rm H}$ values equivalent to A$_{\rm V}$ $\approx$ 6 mag.
Averaging the unabsorbed fluxes from the different fits gives 
L$_{x}$(0.5-7 keV) = 8.1 $\times$10$^{30}$(d$_{pc}$/700)$^{2}$ ergs s$^{-1}$. 
At the nominal cloud distance this L$_{x}$ is slightly less (0.2 dex) than 
CXO 4 in low state, but their distances are unknown.

The observed X-ray flux of this source determined from
the 1T apec Chandra spectral fit is a factor of $\approx$1.4$\times$ greater
 than measured in fits of the XMM-Newton spectra obtained
11 months earlier (Table 3). Thus, the source may be
variable but the flux difference is of low significance
given the measurement uncertainties of
$\pm$35\% for Chandra and $\pm$15\% for XMM-Newton.

\section{Discussion}

\subsection{New YSO Candidates}

The new X-ray data in combination
with 2MASS and VVV nIR colors and Gaia DR3 parallax distances 
reveal new YSO candidates in the cloud.
Five 2MASS and four VVV nIR sources that are matched to X-ray sources 
have nIR color excesses as usually associated with cTTS. 
However, many X-ray sources lack complete JHK$_{s}$ photometry
so the identification of nIR excess sources is not complete.

There are also 27 X-ray sources with Gaia distances 
of 500 - 1000 pc, consistent with previous estimates of
the cloud distance (Sec. 1). They are identified as 
in-cloud (i) sources in Tables 1 and 2. 
These sources could be field stars or YSOs
associated with the cloud. They are prime candidates 
for sensitive follow-up observations at other wavelengths 
to search for evidence of youth such as H$\alpha$ emission,
lithium absorption, and IR color excesses.
There may of course be other X-ray YSO candidates in 
the sample for which Gaia distances are lacking. These 
include those sources identified as variable and 
those whose positions are projected onto the darkest
part of the cloud and identified as on-cloud (o) sources in Tables 1 and 2.

\subsection{What Are the Bright Anonymous X-ray Sources?}

The Chandra sources CXO 4 (= XMM 16) and CXO 21 (= XMM 34) 
are the brightest in terms of count rates but are very faint
(or undetected) in the optical and IR and their
distances are not known. They could be faint stars with 
high levels of X-ray emission or background galaxies.

For comparison with the X-ray properties of galaxies
we searched the Chandra galaxy catalog (Kim et al. 2023)
for known galaxies within a 50$^{\circ}$ radius of the 
X-ray positions. This large search radius returned 309 galaxies,
but none lie within a 18$^{\circ}$ radius.
They have a mean (median) unabsorbed flux of
F$_{x,unabs}$(0.5-7 keV) =
2.2 (0.5) $\times$ 10$^{-14}$ ergs cm$^{-2}$ s$^{-1}$.
WISE colors are available for 270 galaxies and
their mean is W1$-$W2 = 0.30$\pm$0.26 (1$\sigma$)
with a median = 0.27 mag.

CXO 4 has an average flux 
F$_{x,unabs}$(0.5-7 keV) =
34$\pm$10 $\times$ 10$^{-14}$ ergs cm$^{-2}$ s$^{-1}$.
If it is a galaxy its flux would rank it among the 
brightest 1\% of the Chandra sample and thus quite exceptional. 
A TTS at the nominal cloud distance of 700 pc would be optically
faint and a weak-lined TTS (wTTS) could elude detection 
in H$\alpha$ surveys (e.g. Schwartz 1977). But a cTTS 
or wTTS would be brighter in the nIR than the J = 18.5 mag VVV
source located 2$''$ from CXO 4. Using the 
log (L$_{x}$/L$_{*}$) = $-$3.39$\pm$0.06 correlation
for wTTS from Telleschi et al. (2007a) and log L$_{x}$ = 31.15 ergs s$^{-1}$
(low state, at d = 700 pc) gives L$_{*}$ $\approx$ 9 L$_{\odot}$,
equivalent to M$_{V}$ $\approx$ 2.3 mag. 
Adopting A$_{\rm V}$ $\approx$ 7 mag from the X-ray spectral fits 
yields an apparent magnitude V $\approx$ 18.6. Using nIR colors
from Pecault \& Mamajek (2013) for 5 - 30 Myr old stars and
assuming a $\approx$K0-type wTTS as a representative example gives 
J $\approx$ 12 mag, corrected for an extinction 
A$_{\rm J}$ = 0.26A$_{\rm V}$ = 1.8 mag.
No object this bright at J is present in the 2MASS or VVV catalogs
within 3$''$ of the CXO 4 position, so a late-type wTTS at $\sim$700 pc
is effectively ruled out, and likewise for a cTTS.
The possibility of a compact object is discussed below.

CXO 21 also has a high X-ray flux and a
IR counterpart with VVV nIR colors
(H$-$K$_{s}$, J$-$H) = (1.01, 1.12) and  unWISE
color W1$-$W2 = 1.16 mag. Compared to the Chandra sample
its W1$-$W2 color is much too large for a galaxy.
A similar conclusion is reached by comparing its
nIR colors with the galaxy colors of 
Aaronson 
(1977)\footnote{https://ned.ipac.caltech.edu/level5/Sept04/Aaronson/Aaronson4\_3.html}.
But the IR colors of CXO 21 are consistent with a
class I or II YSO (Fig. 10 of Koenig et al. 2012)
and, as noted above, its IR SED slope is consistent with a class II YSO.
However, it is $\approx$5 mag fainter at K$_{s}$ than the
known YSOs Sz 136 and Sz 137 (Fig. 3). If it is a YSO
then its faint K$_{s}$ emission seems to imply
a more distant object behind the cloud.

Based on the above discussion, CXO 21 (= XMM 34) may be an
optically-faint class I/II YSO. But a YSO classification
for CXO 4 (= XMM 16) is by no means established given the lack of 
sufficient data to determine its IR SED.
We thus mention a more speculative possibility.
Reipurth et al. (1997) suggested that star formation in 
Norma may have been triggered by a supernova (SN) explosion 
in the W-NW part of Sa 187, the general area where the two 
bright X-ray sources are seen in projection (Fig. 1).
A SN could leave behind a neutron star (NS) remnant
that would be faint in the optical and IR and possibly 
X-ray bright, as is CXO 4. 
The X-ray spectra of NS are characterized by a
soft thermal blackbody (bb) at
kT$_{\rm bb}$ $\leq$ 0.5 keV and a hotter bb at
kT$_{\rm bb}$ $\approx$ 1-2 keV, or a PL component above 3-4 keV,
may be present (Kaspi \& Boloborodov 2017; 
Borghese \& Esposito 2023; Walter et al. 1996).
X-ray novae can show similar spectra during outbursts
resulting from mass transfer onto a white dwarf (Cannizzo 2000).
The X-ray spectrum of CXO 4 (and CXO 21) does have a hard spectral 
component above 3 keV but line-like features in the 
lightly-binned spectra (Fig. 5 bottom) are not expected for
featureless bb emission. 
Thus, a NS classification seems quite unlikely based 
on the observed spectral properties.
In order to firmly identify and characterize counterparts 
of the two bright mysterious X-ray sources additional sensitive 
multiwavelength observations are clearly needed.

\section{Summary}

We have presented results of the first pointed X-ray observations of
the Norma Dark Cloud. Chandra and XMM-Newton observed the western
filament Sa 187 and detected 121 unique X-ray sources.
Only 7 sources have Gaia DR3 parallax distances $<$500 pc identifying 
them as foreground objects. We find 27 X-ray sources 
with Gaia distances of 500 - 1000 pc, making them probable cloud members 
worthy of multi-wavelength follow-up observations to determine if
they are YSOs. We identify five 2MASS sources and four VVV sources
with nIR excesses (marginal in some cases), as usually associated 
with cTTS. But many X-ray sources lack JHK$_{s}$ photometry so the 
identification of nIR excess sources remains incomplete. 
No X-ray sources were matched to a cataloged galaxy or QSO.
Most faint sources without cataloged counterparts or Gaia distances
are likely background objects but some may be unrecognized heavily-obscured 
low mass objects in the cloud.

The H$\alpha$ emission line stars Sz 136 and Sz 137 were detected
and their Gaia distances of 861-880 pc provide a valuable
calibration of the cloud distance. The A2e star HD 328329 was detected and
its Gaia distance of 538 pc and low X-ray absorption clearly place it on
the near side of the cloud. The peculiar FU Ori-like star V346 Nor and
nebulous star Re 13 were not detected.

An intriguing result is that the two brightest Chandra X-ray sources 
CXO 4 and CXO 21 have very faint optical and IR counterparts without
known distances and the counterpart of CXO 4 is ambiguous. 
The type of objects giving rise to their
X-ray emission is uncertain. Both sources show signs of magnetic
activity including a hard
spectral component above 3 keV arising either from hot thermal 
plasma or nonthermal emission. Sensitive multiwavelength follow-up 
observations of these two enigmatic X-ray sources will be needed to 
determine their nature.

\clearpage

\begin{acknowledgments}
Support for this work was provided by  {\em Chandra} award
number GO1-22008X issued by the {\em Chandra} X-ray Center, which is operated by
the Smithsonian Astrophysical Observatory (SAO) for and on behalf of NASA.
Support was also provided by NASA/GSFC award 80NSSC22K0689. Observations
were obtained with XMM-Newton, an ESA science mission directly funded 
by ESA member states and the USA (NASA).
This research has made use of the NASA/IPAC Infrared Science Archive (IRSA), 
which is funded by NASA and operated by the California Institute of Technology.
This work has made use of data products from the Two Micron All Sky Survey 
(2MASS; Skrutskie et al. 2003; 2006), 
which is a joint project of the University of Massachusetts and the Infrared Processing and 
Analysis Center (IPAC)/California Institute of Technology, funded by NASA and the NSF, 
and data products from the 
Wide-field Infrared Survey Explorer (WISE; Wright et al. 2010b; 2019)  which is a joint project 
of the University of California, Los Angeles, and the Jet Propulsion Laboratory/California 
Institute of Technology, funded by NASA. 
This work has made use of data from the unWISE Catalog (Schlafly et al. 2019; unWISE Team 2021).
This work has made use of data from the European Space Agency (ESA) {\it Gaia} mission,
(\url{https://www.cosmos.esa.int/gaia}; IRSA 2022), 
processed by the {\it Gaia} Data Processing and Analysis Consortium. \\

\noindent This paper employs a list of {\em Chandra} datasets obtained by the
Chandra X-ray Observatory contained in
~\dataset[DOI: 10.25574/cdc.341]{https://doi.org/10.25574/cdc.341}.



\end{acknowledgments}

\vspace{5mm}
\facilities{CXO; IRSA; WISE; XMM}

\vspace{5mm}
\software{XSPEC (Arnaud 1996),
          CIAO (Fruscione et al. 2006),
          SAS (Gabriel~ et al. 2004)}

\clearpage
\startlongtable
\begin{deluxetable}{lllllllllcl}
\tabletypesize{\scriptsize}
\tablecaption{Chandra X-ray Sources in Norma}
\tablehead{
           \colhead{CXO}           &
           \colhead{XMM}           &
           \colhead{R.A.}          &
           \colhead{Decl.}         &
           \colhead{Net Cts}       &
           \colhead{Rate}          &
           \colhead{E$_{50}$}      &
           \colhead{H.R.}          &
           \colhead{nIR}           &
           \colhead{Gaia offset}          &
           \colhead{Notes}         \\
           \colhead{nr.}           &
           \colhead{nr.}           &
           \colhead{(J2000)}       &
           \colhead{(J2000)}       &
           \colhead{(c)}           &
           \colhead{(c ks$^{-1}$)} &
           \colhead{(keV)}         &
           \colhead{}              &
           \colhead{(J2000)}       &
           \colhead{($''$)}        &
           \colhead{   }          }
\startdata
1 f    & 11  & 16 31 53.45& -44 53 08.0 &   65$\pm$9   & 1.35$\pm$0.19 & 1.62  & 0.34  & 2MA 16315336-4453078 (0.9) & 0.52 & 1,3,5   \\  
2      & 12  & 16 31 54.03& -44 48 55.4 &   57$\pm$9   & 1.27$\pm$0.20 & 3.38  & 0.86  & 2MA 16315410-4448584 (3.1) & ...  & 2,4  \\
3 f    & 15  & 16 32 01.98& -44 49 45.5 &   36$\pm$7   & 0.75$\pm$0.14 & 2.50  & 0.67  &                            & 2.06 &         \\
4      & 16  & 16 32 02.25& -44 53 48.5 & 587$\pm$25   & 11.3$\pm$0.48 & 2.75  & 0.72  &                            & 2.26 & 6        \\
5      & 17  & 16 32 04.52& -44 53 59.5 &   57$\pm$8   & 1.14$\pm$0.17 & 3.20  & 0.84  &                            & ...  &         \\
6      & ... & 16 32 04.74& -44 56 47.7 &   11$\pm$4   & 0.22$\pm$0.08 & 2.38  & 0.66  &                            & 0.38 & 5         \\
7      & 19  & 16 32 07.82& -44 58 54.7 &   93$\pm$11  & 1.75$\pm$0.20 & 3.75  & 0.93  &                            & ...  &         \\
8      & 22  & 16 32 10.92& -44 50 03.4 &   51$\pm$8   & 1.19$\pm$0.19 & 2.25  & 0.62  &                            & ...  &         \\
9 i    & ... & 16 32 11.70& -44 57 16.8 &    9$\pm$4   & 0.17$\pm$0.07 & 2.50  & 0.62  & 2MA 16321146-4457150 (3.1) & 0.70 & 3,4,7,9        \\
10     & 25  & 16 32 12.70& -44 51 52.4 &   53$\pm$8   & 1.02$\pm$0.16 & 2.88  & 0.76  & 2MA 16321268-4451542 (1.9) & 0.91 & 2,3,4,9        \\
11 i   & 26  & 16 32 12.75& -44 55 11.2 &   17$\pm$4   & 0.47$\pm$0.13 & 1.75  & 0.44  & 2MA 16321268-4455119 (1.0) & 1.09 & 3       \\
12     & 27  & 16 32 13.11& -44 57 23.8 &   22$\pm$5   & 0.40$\pm$0.09 & 2.38  & 0.51  & 2MA 16321311-4457241 (0.3) & 0.33 & 1,3        \\
13     & 28  & 16 32 13.28& -44 59 03.9 &   21$\pm$5   & 0.37$\pm$0.10 & 1.63  & 0.25  &                            & 1.31 & 2,7       \\
14     & 29  & 16 32 13.56& -44 59 32.7 &   22$\pm$6   & 0.40$\pm$0.10 & 3.00  & 0.82  &                            & ...  &         \\
15     & 30  & 16 32 14.84& -44 56 35.4 &   99$\pm$10  & 1.77$\pm$0.18 & 3.13  & 0.75  &                            & ...  &        \\
16     & 31  & 16 32 15.71& -44 49 18.9 &   57$\pm$9   & 1.08$\pm$0.16 & 2.75  & 0.71  & 2MA 16321584-4449176 (1.9) & 1.77 & 2    \\
17     & ... & 16 32 16.62& -45 01 25.2 &   14$\pm$5   & 0.28$\pm$0.09 & 1.75  & 0.35  & 2MA 16321674-4501231 (2.5) & 1.10 & 9          \\
18     & ... & 16 32 20.21& -44 54 02.9 &   11$\pm$4   & 0.19$\pm$0.06 & 3.13  & 0.69  &                            & ...  &       \\
19     & ... & 16 32 21.68& -44 51 49.2 &   12$\pm$4   & 0.21$\pm$0.07 & 2.50  & 0.90  &                            & 1.57 & 7       \\
20     & ... & 16 32 22.79& -44 56 13.3 &    7$\pm$3   & 0.13$\pm$0.05 & 4.62  & 1.00  &                            & ...  & 7    \\
21     & 34  & 16 32 23.64& -44 48 38.6 &  305$\pm$18  & 6.21$\pm$0.38 & 2.90  & 0.74  & VVV 16322361-4448385 (0.2) & ...  & 8        \\
22 o   & ... & 16 32 25.54& -44 56 47.8 &   17$\pm$4   & 0.34$\pm$0.08 & 4.25  & 1.00  &                            & ...  &         \\
23     & ... & 16 32 26.69& -44 57 40.4 &   12$\pm$4   & 0.26$\pm$0.08 & 2.63  & 0.69  &                            & 2.47 &          \\
24     & ... & 16 32 27.56& -44 57 39.2 &   43$\pm$7   & 0.75$\pm$0.12  & 4.00 & 0.87  & 2MA 16322768-4457381 (1.6) & 1.37 & 2         \\
25     & ... & 16 32 28.74& -44 58 10.4 &   11$\pm$4   & 0.18$\pm$0.06  & 1.75 & 0.32  &                            & ...  & 7        \\
26     & 36  & 16 32 28.90& -44 58 19.7 &   19$\pm$5   & 0.33$\pm$0.08  & 3.50 & 0.86  & 2MA 16322904-4458219 (2.7) & 2.75 & 2         \\
27     & 35  & 16 32 29.03& -44 57 38.5 &   49$\pm$7   & 0.82$\pm$0.12  & 3.63 & 0.82  &                            & ...  &          \\
28 i   & 37  & 16 32 30.92& -44 51 50.7 &   18$\pm$5   & 0.34$\pm$0.09  & 1.25 & 0.35  & 2MA 16323090-4451520 (1.4) & 1.48 & 1,9          \\
29 o   & ... & 16 32 31.36& -44 56 18.6 &   11$\pm$4   & 0.19$\pm$0.06  & 4.37 & 0.84  &                            & ...  &          \\
30 o   & ... & 16 32 31.95& -44 53 54.0 &    8$\pm$3   & 0.18$\pm$0.07  & 4.12 & 0.88  &                            & ...  & 7       \\
31 i   & 39  & 16 32 32.67& -44 56 38.5 &    8$\pm$3   & 0.13$\pm$0.05  & 1.13 & 0.00  & 2MA 16323267-4456389 (0.4) & 0.58 & 3,7,9,10  \\
32     & ... & 16 32 32.91& -44 56 44.1 &    7$\pm$3   & 0.13$\pm$0.05  & 4.24 & 0.75  &                            & ...  & 7         \\
33     & 40  & 16 32 35.50& -45 01 49.7 &   19$\pm$5   & 0.44$\pm$0.12  & 2.88 & 0.90  & VVV 16323568-4501516 (2.8) & ...  & 7,8        \\
34     & 41  & 16 32 36.73& -45 02 30.2 &   25$\pm$6   & 0.52$\pm$0.12  & 3.50 & 0.67  & VVV 16323668-4502309 (0.9) & ...  & 8        \\
35 o   & ... & 16 32 37.86& -44 55 13.3 &   18$\pm$4   & 0.36$\pm$0.09  & 2.99 & 0.68  & 2MA 16323769-4455117 (2.3) & ...  & 7         \\
36 o   & 44  & 16 32 38.90& -44 54 42.5 &   14$\pm$4   & 0.26$\pm$0.07  & 4.62 & 0.94  &                            & ...  &          \\
37     & ... & 16 32 40.15& -45 01 45.3 &   23$\pm$6   & 0.46$\pm$0.11  & 3.13 & 0.83  &                            & 2.95 & 3,7,9        \\
38 i   & 47  & 16 32 40.83& -44 59 28.8 &   21$\pm$5   & 0.38$\pm$0.10  & 1.63 & 0.31  & 2MA 16324088-4459301 (1.5) & 0.35 & 1,3,5,9         \\
39     & 46  & 16 32 40.94& -44 51 45.3 &   10$\pm$4   & 0.19$\pm$0.07  & 3.63 & 1.00  &                            & ...  & 7        \\
40     & 51  & 16 32 41.02& -44 49 01.7 &   31$\pm$7   & 0.64$\pm$0.14  & 3.13 & 0.77  &                            & 2.58 & 9      \\
41     & 49  & 16 32 41.08& -44 48 22.3 &   22$\pm$6   & 0.55$\pm$0.15  & 4.00 & 0.94  &                            & 2.37 & 7       \\
42     & 50  & 16 32 41.30& -44 50 40.0 &   41$\pm$7   & 0.79$\pm$0.14  & 3.38 & 0.84  &                            & 2.22 & 3,9       \\
43 f   & 52  & 16 32 41.47& -45 00 37.1 &   26$\pm$6   & 0.48$\pm$0.11  & 1.38 & 0.30  & 2MA 16324138-4500362 (1.3) & 1.17 & 1,3,9          \\
44     & 53  & 16 32 43.26& -44 52 08.8 &   33$\pm$6   & 0.59$\pm$0.11  & 3.38 & 0.76  & 2MA 16324317-4452061 (2.8) & 1.33 & 3,9        \\
45 o   & 54  & 16 32 44.26& -44 53 43.1 &   68$\pm$8   & 1.21$\pm$0.15  & 3.50 & 0.86  &                            & ...  &        \\
46 o   & 57  & 16 32 46.84& -45 01 17.5 &   17$\pm$5   & 0.39$\pm$0.12  & 3.50 & 0.96  &                            & ...  & 7        \\
47 o   & ... & 16 32 47.33& -44 56 51.7 &    9$\pm$3   & 0.22$\pm$0.08  & 1.25 & 0.25  & 2MA 16324736-4456522 (0.7) & 0.76 & 2,8         \\
48     & 59  & 16 32 47.43& -44 50 51.7 &   30$\pm$7   & 0.56$\pm$0.12  & 3.00 & 0.63  &                            & 1.87 & 7        \\
49 o   & ... & 16 32 48.04& -44 55 45.8 &    8$\pm$3   & 0.16$\pm$0.06  & 2.13 & 0.52  &                            & ...  & 7         \\
50 i,o & 61  & 16 32 49.25& -44 56 45.1 &   73$\pm$9   & 1.68$\pm$0.20  & 3.00 & 0.68  & 2MA 16324904-4456444 (2.3) & 2.19 &          \\
51 o   & 62  & 16 32 49.74& -45 01 01.3 &   39$\pm$7   & 0.76$\pm$0.14  & 1.63 & 0.23  & 2MA 16324980-4501020 (1.0) & 1.09 & 1,3          \\
52 i   & 65  & 16 32 50.73& -44 50 43.0 &   32$\pm$6   & 0.58$\pm$0.12  & 1.50 & 0.19  & 2MA 16325072-4450441 (1.2) & 0.83 & 1,3,9          \\
53 i   & 64  & 16 32 50.84& -44 49 36.1 &   12$\pm$4   & 0.48$\pm$0.12  & 1.75 & 0.29  & 2MA 16325066-4449369 (2.1) & 1.89 & 3,7,9          \\
54 o   & 66  & 16 32 51.35& -44 54 44.8 &   12$\pm$4   & 0.22$\pm$0.07  & 1.75 & 0.39  & 2MA 16325134-4454453 (0.5) & 0.67 & 2,5,7,9          \\
55     & ... & 16 32 52.44& -44 50 23.0 &   19$\pm$5   & 0.36$\pm$0.10  & 4.38 & 0.98  &                            & ...  &           \\
56 o   & ... & 16 32 56.85& -44 59 23.4 &   16$\pm$5   & 0.32$\pm$0.09  & 4.12 & 0.80  &                            & ...  & 7         \\
57 i,o & ... & 16 32 57.68& -44 55 28.2 &   21$\pm$6   & 0.40$\pm$0.11  & 1.90 & 0.46  & 2MA 16325769-4455287 (0.5) & 0.67 & 3,9         \\
58 o   & 69  & 16 32 58.15& -44 56 57.9 &   33$\pm$7   & 0.62$\pm$0.13  & 4.00 & 0.86  &                            & ...  &         \\
59     & 70  & 16 32 58.33& -44 54 26.9 &   20$\pm$5   & 0.40$\pm$0.10  & 3.13 & 0.84  & 2MA 16325812-4454249 (2.9) & 2.90 & 7        \\
60 o   & ... & 16 33 00.50& -45 01 24.5 &   14$\pm$5   & 0.30$\pm$0.10  & 2.60 & 0.48  & VVV 16330054-4501264 (2.0) & ...  & 3,7,8         \\
61 o   & 74  & 16 33 00.82& -44 57 51.8 &   31$\pm$6   & 0.60$\pm$0.13  & 1.75 & 0.42  & 2MA 16330075-4457515 (0.7) & 0.60 & 9         \\
62 i,o & ... & 16 33 03.38& -44 58 04.5 &   13$\pm$5   & 0.29$\pm$0.09  & 2.30 & 0.54  & 2MA 16330342-4458070 (2.6) & 2.67 & 2,7         \\
63 o   & ... & 16 33 03.55& -44 57 29.4 &   23$\pm$6   & 0.57$\pm$0.15  & 3.38 & 0.81  &                            & ...  & 5,7       \\
64 i,o & 78  & 16 33 04.53& -44 57 16.8 &   24$\pm$6   & 0.47$\pm$0.12  & 1.63 & 0.30  & 2MA 16330448-4457181 (1.4) & 1.55 & 3,6,8,14,11  \\
65     & ... & 16 33 05.07& -44 53 51.4 &   12$\pm$5   & 0.23$\pm$0.08  & 2.75 & 0.59  &                            & 0.28 &       \\
66 i   & 79  & 16 33 05.30& -44 54 25.1 &   21$\pm$6   & 0.40$\pm$0.11  & 1.88 & 0.49  & 2MA 16330519-4454254 (1.2) & 1.09 & 2,3,7,9         \\
67     & 80  & 16 33 05.92& -44 50 02.3 &   30$\pm$7   & 0.60$\pm$0.14  & 3.13 & 0.57  & 2MA 16330579-4450035 (1.9) & 1.28 & 2,7,8,9          \\
68 o   & ... & 16 33 06.86& -44 59 47.2 &   33$\pm$7   & 0.72$\pm$0.15  & 2.38 & 0.68  &                            & ...  & 5         \\
69     & ... & 16 33 07.04& -44 53 01.2 &   14$\pm$5   & 0.26$\pm$0.09  & 1.70 & 0.46  &                            & ...  & 7        \\
70 o   & ... & 16 33 08.10& -44 56 29.9 &   32$\pm$7   & 0.71$\pm$0.15  & 4.12 & 0.97  &                            & ...  &         \\
71 o   & ... & 16 33 09.84& -44 57 12.9 &   27$\pm$7   & 0.55$\pm$0.14  & 2.75 & 0.71  & VVV 16330969-4457140 (2.0) & ...  & 7,8      \\
72     & 83  & 16 33 10.96& -44 50 17.5 &   33$\pm$7   & 0.70$\pm$0.15  & 3.62 & 0.97  &                            & ...  &          \\
73 i,o & 84  & 16 33 12.00& -44 56 13.9 &   56$\pm$10  & 1.15$\pm$0.20  & 2.50 & 0.74  & 2MA 16331205-4456141 (0.6) & 0.78 & 3,5,6,8,12    \\
74 o   & 86  & 16 33 15.17& -44 58 26.0 &   17$\pm$5   & 0.65$\pm$0.20  & 1.80 & 0.21  & 2MA 16331497-4458245 (2.6) & ...  & 4,7        \\
75     & ... & 16 33 15.30& -44 51 14.6 &   13$\pm$5   & 0.27$\pm$0.10  & 2.38 & 0.71  & 2MA 16331533-4451136 (1.0) & 1.02 & 2,3,7,9        \\
\enddata
\tablecomments{
In column 1, f = foreground source (Gaia parallax distance d $<$ 500 pc),
i = possible in-cloud source - there is at least one Gaia DR3 object within 3$''$
at d = 500 - 1000 pc,
o = on-cloud source, X-ray position is projected onto the darkest part of the cloud (Fig. 1).
J2000 X-ray position (R.A., Decl.), Net Counts, and Rate are computed using events in 
the 0.5-7 keV range inside 3$\sigma$ source position ellipses as determined by CIAO {\em wavdetect} 
applied to the merged observations. Rate = (Net Cts)/Exptime where Exptime is the effective
exposure time for each source based on the merged data exposure map and is slightly different 
for each source. E$_{50}$ is the median event energy of 0.5-7 keV events and is 
background subtracted using the method of Hong et al. (2004). 
Hardness Ratio H.R. = counts(2-7 keV)/counts(0.5-7 keV) and the counts in each energy
band are background subtracted.
Possible 2MASS nIR counterparts lying within 3$''$ of the X-ray position are identified as 2MA with the 
CXO-2MASS position offset (arcsec) in parentheses after 2MASS Id. 
VVV sources may also be present within 3$''$ of the X-ray positions but are only                            
listed (with position offsets in parentheses) if they have a nIR excess and no candidate 2MASS
counterpart was found.
The offset between the Gaia DR3 and Chandra positions is given for any Gaia object 
whose position lies inside the Chandra 3$\sigma$ position errror ellipse. If multiple
Gaia objects fall within the ellipse the offset of the object closest to the 
Chandra position is given.
\\
Notes. \\
       (1) There is a HST GSC class=0 (star) object within 3$''$ of the X-ray position.\\
       (2) There is a HST GSC class=3 (non-star) object within 3$''$ of the Chandra position.\\
       (3) There is a WISE ALL-Sky Survey IR object within 3$''$ of the Chandra position. \\
       (4) The 2MASS identification is uncertain due to an offset $>$3$''$ or multiple 
           2MASS sources near the X-ray position. \\
       (5) Source count rate varied significantly during one or more observations 
           (CIAO {\em glvary} VARINDEX $\geq$ 7). \\
       (6) Source mean count rate differed significantly between two or more observations. \\
       (7) The source was only detected in the merged observations. \\
       (8) The identified 2MASS or VVV source has a near-IR color excess.  \\
       (9) There is more than one Gaia DR3 object within 3$''$ of the Chandra position. \\
       (10) Counterpart is the A2e star HD 328329. \\
       (11) Counterpart is the emission line star Sz 136. \\
       (12) Counterpart is the emission line star Sz 137. 
}
\end{deluxetable}

\clearpage
\startlongtable
\begin{deluxetable}{llllllclll}
\tabletypesize{\scriptsize}            
\tablewidth{0pt}
\tablecaption{XMM-Newton X-ray Sources in Chandra ACIS-I Norma Field of View }
\tablehead{
           \colhead{XMM}           &
           \colhead{CXO}           &
           \colhead{R.A.}          &
           \colhead{Decl.}         &
           \colhead{Counts}        &
           \colhead{Counts}        &
           \colhead{H.R.(pn)}      &
           \colhead{nIR}           &
           \colhead{Gaia}         &
           \colhead{Notes}           \\
           \colhead{nr.}          &
           \colhead{nr.}          &
           \colhead{}             &
           \colhead{}             &
           \colhead{pn}           &
           \colhead{MOS1$+$2}     &
           \colhead{}             &
           \colhead{}             &
           \colhead{offset}       &
           \colhead{}             \\
           \colhead{}              &
           \colhead{}              &
           \colhead{(J2000)}       &
           \colhead{(J2000)}       &
           \colhead{(cts)}         &
           \colhead{(cts)}         &
           \colhead{}              &
           \colhead{(J2000)}       &
           \colhead{($''$)}        &
           \colhead{   }          }
\startdata
1          & ... & 16 31 37.68 & -44 59 59.4    & 68$\pm$14    &  14$\pm$9    & 0.40  & 2MA 16313788-4459574 (2.9) & 0.99  & 4,6    \\
2          & ... & 16 31 40.46 & -44 55 53.6    & 40$\pm$17    &  48$\pm$19   & 0.63  &                            & ...   & 5    \\
3          & ... & 16 31 43.89 & -44 56 35.3    & 36$\pm$13    &  44$\pm$18   & 0.64  &                            & ...   &     \\
4          & ... & 16 31 45.37 & -44 55 28.0    & 81$\pm$13    &  20$\pm$15   & 0.00  & ...                        & ...   &     \\
5          & ... & 16 31 46.27 & -44 58 25.5    & 74$\pm$15    &  54$\pm$20   & 0.61  &                            & ...   &     \\
6          & ... & 16 31 46.41 & -44 57 55.3    & 43$\pm$11    &  11$\pm$11   & 0.05  & 2MA 16314642-4457575 (2.2) & 1.60  &    \\
7          & ... & 16 31 49.41 & -45 02 32.3    & 81$\pm$20    &  14$\pm$13   & 0.60  &                            & 2.87* & 5    \\
8          & ... & 16 31 49.44 & -45 00 19.6    & 509$\pm$209  &  249$\pm$167 & 0.82  &                            & ...   & 5     \\
9 i        & ... & 16 31 49.51 & -44 54 41.7    & 66$\pm$13    &  59$\pm$18   & 0.01  & 2MA 16314952-4454412 (0.5) & 0.40  & 1    \\
10         & ... & 16 31 52.81 & -44 55 55.1    & 112$\pm$21   & 102$\pm$32   & 0.56  & 2MA 16315283-4455521 (3.0) & ...   &     \\
11 f       &  1  & 16 31 53.38 & -44 53 07.8    & 1120$\pm$38  & 571$\pm$40   & 0.05  & 2MA 16315336-4453078 (0.1) & 0.36  & 1,3  \\
12         &  2  & 16 31 54.06 & -44 48 53.7    & 113$\pm$17   &  94$\pm$22   & 0.68  & 2MA 16315410-4448584 (4.8) & ...   & 6     \\
13         & ... & 16 31 58.32 & -44 47 37.2    & 41$\pm$12    &  48$\pm$19   & 0.79  & ...                        & ...   &     \\
14 i       & ... & 16 32 00.55 & -44 46 58.7    & 74$\pm$13    &  57$\pm$17   & 0.01  & 2MA 16320054-4446588 (0.2) & 0.30  & 4    \\
15 f       &  3  & 16 32 02.05 & -44 49 44.7    & 78$\pm$15    &  71$\pm$20   & 0.77  & ...                        & 1.02  &     \\
16         &  4  & 16 32 02.28 & -44 53 48.3    & 811$\pm$35   & 309$\pm$46   & 0.56  &                            & 2.24* &     \\
17         &  5  & 16 32 04.46 & -44 54 00.6    & 83$\pm$17    &  70$\pm$19   & 0.90  &                            & ...   &     \\
18         & ... & 16 32 07.68 & -44 55 30.3    & 72$\pm$15    &  18$\pm$13   & 0.29  & ...                        & ...   &     \\
19         &  7  & 16 32 07.85 & -44 58 54.4    & 159$\pm$18   & 128$\pm$24   & 1.00  & ...                        & ...   &     \\
20         & ... & 16 32 08.15 & -44 47 17.3    & 49$\pm$12    &   3$\pm$5    & 0.95  & ...                        & ...   &     \\
21 f       & ... & 16 32 08.84 & -44 53 48.6    & 60$\pm$13    &  38$\pm$15   & 0.00  & 2MA 16320879-4453501 (1.6) & 1.89* & 1,3,4  \\
22         &  8  & 16 32 10.84 & -44 50 05.8    & 91$\pm$16    &  54$\pm$19   & 0.81  & ...                        & ...   &     \\
23 i       & ... & 16 32 12.07 & -45 00 08.1    & 43$\pm$14    &  46$\pm$17   & 0.53  &                            & 1.88* &     \\
24         & ... & 16 32 12.13 & -45 01 15.1    & 78$\pm$16    &  16$\pm$7    & 0.57  &                            & ...   & 5    \\
25         & 10  & 16 32 12.60 & -44 51 52.1    & 139$\pm$18   & 111$\pm$22   & 0.81  & 2MA 16321268-4451542 (2.3) & 0.68  & 3,4   \\
26 i       & 11  & 16 32 12.78 & -44 55 12.1    & 54$\pm$13    &  64$\pm$19   & 0.14  & 2MA 16321268-4455119 (1.0) & 0.95  & 3   \\
27         & 12  & 16 32 12.98 & -44 57 23.4    & 92$\pm$16    &  50$\pm$16   & 0.30  & 2MA 16321311-4457241 (1.6) & 1.63* &     \\
28         & 13  & 16 32 13.15 & -44 59 01.9    & 91$\pm$17    &  48$\pm$17   & 0.31  &                            & 1.97* & 2   \\
29         & 14  & 16 32 13.60 & -44 59 31.0    & 55$\pm$14    &  38$\pm$17   & 0.64  &                            & ...   & 5    \\
30         & 15  & 16 32 14.81 & -44 56 35.5    & 283$\pm$22   & 235$\pm$29   & 0.70  & ...                        & ...   &     \\
31         & 16  & 16 32 15.59 & -44 49 19.1    & 151$\pm$19   &  39$\pm$23   & 0.70  & ...                        & 2.96* &     \\
32 i       & ... & 16 32 16.76 & -44 48 44.7    & 41$\pm$11    &   6$\pm$7    & 0.10  & 2MA 16321672-4448463 (1.8) & 1.04  & 2,4   \\
33 i       & ... & 16 32 20.48 & -44 56 15.1    & 213$\pm$65   &  93$\pm$64   & 0.83  & 2MA 16322067-4456162 (2.5) & 2.28* & 1,4,5   \\
34         & 21  & 16 32 23.66 & -44 48 38.7    & 575$\pm$30   & 204$\pm$18   & 0.72  & VVV 16322361-4448385 (0.5) & ...   & 7    \\
35         & 27  & 16 32 28.85 & -44 57 38.2    & 154$\pm$19   & 102$\pm$22   & 0.56  & ...                        & ...   &     \\
36         & 26  & 16 32 28.89 & -44 58 19.7    & 171$\pm$20   & 125$\pm$23   & 0.82  & 2MA 16322904-4458219 (2.8) & 2.81* & 2   \\
37 i       & 28  & 16 32 30.94 & -44 51 51.5    & 176$\pm$18   & 129$\pm$22   & 0.04  & 2MA 16323090-4451520 (0.7) & 0.65  & 1,4   \\
38 i       & ... & 16 32 31.30 & -44 48 45.4    & 37$\pm$12    &  30$\pm$10   & 0.03  &                            & 1.03  & 4    \\
39 i       & 31  & 16 32 32.77 & -44 56 38.3    & 124$\pm$17   &  72$\pm$18   & 0.20  & 2MA 16323267-4456389 (1.2) & 1.36  & 1,3 HD 328329   \\
40         & 33  & 16 32 35.81 & -45 01 50.5    & 80$\pm$16    &  24$\pm$8    & 0.73  & VVV 16323568-4501516 (1.8) & ...   & 7    \\
41         & 34  & 16 32 36.77 & -45 02 30.7    & 52$\pm$14    &  18$\pm$7    & 0.75  & VVV 16323668-4502309 (1.0) & ...   & 7    \\
42 f,o     & ... & 16 32 36.86 & -45 03 06.2    & 48$\pm$13    &  30$\pm$9    & 0.37  & 2MA 16323684-4503045 (1.6) & 0.53  & 2,4   \\
43 f,o     & ... & 16 32 37.52 & -44 52 54.4    & 57$\pm$14    &  25$\pm$13   & 0.06  & 2MA 16323773-4452544 (2.3) & 1.78* & 1,3,4,TYC7866-2460-1  \\
44 o       & 36  & 16 32 38.85 & -44 54 44.2    & 41$\pm$12    &  32$\pm$14   & 0.79  &                            & ...   &     \\
45         & ... & 16 32 39.20 & -44 47 09.3    & 54$\pm$14    &  16$\pm$8    & 0.65  & ...                        & ...   &     \\
46         & 39  & 16 32 40.77 & -44 51 43.6    & 54$\pm$13    &  40$\pm$14   & 0.83  &                            & ...   &     \\
47 i       & 38  & 16 32 41.01 & -44 59 29.0    & 192$\pm$19   & 105$\pm$21   & 0.08  & 2MA 16324088-4459301 (1.7) & 0.86  & 1,3,4  \\
48 o       & ... & 16 32 41.05 & -44 53 07.8    & 30$\pm$12    &  33$\pm$13   & 0.34  & 2MA 16324124-4453055 (3.0) & ...   &     \\
49         & 41  & 16 32 41.12 & -44 48 25.2    & 50$\pm$14    &  17$\pm$7    & 0.73  &                            & ...   &     \\
50         & 42  & 16 32 41.22 & -44 50 40.6    & 120$\pm$17   & 104$\pm$21   & 0.65  & ...                        & ...   &     \\
51         & 40  & 16 32 41.23 & -44 49 02.8    & 46$\pm$13    &  18$\pm$7    & 0.87  &                            & 0.57  &     \\
52 f       & 43  & 16 32 41.39 & -45 00 36.3    & 267$\pm$21   & 177$\pm$25   & 0.06  & 2MA 16324138-4500362 (0.1) & 0.12  & 1,3,4,5 \\
53         & 44  & 16 32 43.13 & -44 52 08.2    & 67$\pm$14    &  46$\pm$16   & 0.67  & 2MA 16324317-4452061 (2.2) & 0.29  & 2,3,4 \\
54 o       & 45  & 16 32 44.38 & -44 53 42.9    & 118$\pm$16   &  99$\pm$21   & 0.83  &                            & ...   &     \\
55 i,o     & ... & 16 32 45.16 & -44 56 33.7    & 42$\pm$25    &  65$\pm$35   & 0.46  & 2MA 16324515-4456337 (0.1) & 0.18  & 1,5   \\
56         & ... & 16 32 45.71 & -44 59 53.1    & 37$\pm$11    &  32$\pm$19   & 0.31  & 2MA 16324571-4459557 (2.7) & 2.71* &     \\
57         & 46  & 16 32 47.23 & -45 01 17.7    & 63$\pm$14    &  28$\pm$15   & 0.68  &                            & ...   &     \\
58 i       & ... & 16 32 47.25 & -44 52 05.5    & 54$\pm$13    &  41$\pm$15   & 0.22  & 2MA 16324697-4452057 (2.9) & 2.27* & 2,4    \\
59         & 48  & 16 32 47.28 & -44 50 51.3    & 144$\pm$17   & 141$\pm$24   & 0.57  &                            & 1.35  &     \\
60 i       & ... & 16 32 48.57 & -44 47 03.6    & 79$\pm$13    &  18$\pm$7    & 0.00  & 2MA 16324854-4447040 (0.6) & 0.59  & 2    \\
61 i,o     & 50  & 16 32 49.17 & -44 56 44.6    & 120$\pm$22   &  218$\pm$28  & 0.71  & 2MA 16324904-4456444 (1.4) & 1.30  & 2,3,5   \\
62         & 51  & 16 32 49.83 & -45 01 00.5    & 179$\pm$18   &  122$\pm$22  & 0.25  & 2MA 16324980-4501020 (1.6) & 1.65* & 1,3   \\
63 i       & ... & 16 32 50.09 & -44 48 56.8    & 32$\pm$11    &  32$\pm$9    & 0.35  & 2MA 16324988-4448566 (2.2) & 2.08* & 2,3,4   \\
64 i       & 53  & 16 32 50.50 & -44 49 36.5    & 75$\pm$13    &  32$\pm$9    & 0.11  & 2MA 16325066-4449369 (1.8) & 1.81* & 2,3,4   \\
65 i       & 52  & 16 32 50.69 & -44 50 43.5    & 171$\pm$32   & 388$\pm$35   & 0.44  & 2MA 16325072-4450441 (0.7) & 0.37  & 1,3,4,5   \\
66 o       & 54  & 16 32 51.43 & -44 54 46.5    & 48$\pm$12    &  10$\pm$11   & 0.04  & 2MA 16325134-4454453 (1.4) & 1.30  & 2,    \\
67 i       & ... & 16 32 52.86 & -45 03 51.0    & 627$\pm$31   &  91$\pm$14   & 0.51  & 2MA 16325287-4503493 (1.7) & 0.59  & 2,3,4    \\
68         & ... & 16 32 53.98 & -44 52 45.2    & 60$\pm$14    &  51$\pm$17   & 0.58  &                            & ...   &     \\
69 o       & 58  & 16 32 58.36 & -44 56 58.7    & 38$\pm$16    &  49$\pm$17   & 1.00  &                            & ...   & 5    \\
70         & 59  & 16 32 58.51 & -44 54 26.0    & 70$\pm$14    &  55$\pm$18   & 0.72  &                            & 1.67* &     \\
71 i,o     & ... & 16 32 58.67 & -45 00 44.3    & 52$\pm$12    &  32$\pm$17   & 0.15  & 2MA 16325860-4500446 (0.7) & 0.99  & 1,3 \\
72         & ... & 16 32 58.72 & -45 03 07.6    & 39$\pm$12    &  32$\pm$15   & 0.16  & 2MA 16325871-4503060 (1.6) & 1.41  & 2   \\
73 i       & ... & 16 33 00.66 & -44 51 44.8    & 48$\pm$13    &  32$\pm$15   & 0.06  & 2MA 16330044-4451456 (2.4) & 1.46  & 2,4   \\
74 o       & 61  & 16 33 00.83 & -44 57 50.7    & 99$\pm$16    &  52$\pm$16   & 0.27  & 2MA 16330075-4457515 (1.1) & 1.18  & 5    \\
75 o       & ... & 16 33 03.55 & -44 57 47.4    & 43$\pm$12    &  43$\pm$17   & 0.15  & 2MA 16330376-4457481 (2.4) & 2.02* & 1,3,4  \\
76         & ... & 16 33 03.87 & -44 46 51.9    & 68$\pm$14    &  29$\pm$9    & 0.63  &                            & 2.26* &     \\
77 o       & ... & 16 33 03.93 & -44 58 07.7    & 35$\pm$13    &  22$\pm$14   & 0.36  & ...                        & ...   & 5    \\
78 i,o     & 64  & 16 33 04.43 & -44 57 18.6    & 192$\pm$18   & 180$\pm$26   & 0.31  & 2MA 16330448-4457181 (0.8) & 0.76  & 1,3,Sz136  \\
79 i       & 66  & 16 33 05.37 & -44 54 26.2    & 25$\pm$10    &  48$\pm$17   & 0.00  & 2MA 16330519-4454254 (2.0) & 1.91* & 2,3,4    \\
80         & 67  & 16 33 05.73 & -44 50 06.3    & 72$\pm$14    &  17$\pm$8    & 0.52  & 2MA 16330579-4450035 (2.8) & 2.93* & 2    \\
81 o       & ... & 16 33 08.38 & -44 59 48.8    & 217$\pm$24   & 393$\pm$36   & 0.66  & 2MA 16330838-4459485 (0.2) & ...   & 5    \\
82 f       & ... & 16 33 09.96 & -44 54 14.4    & 87$\pm$13    &  45$\pm$17   & 0.01  & 2MA 16331001-4454122 (2.2) & 1.55  & 1,3,TYC7866-2378-1 \\
83         & 72  & 16 33 11.30 & -44 50 17.8    & 55$\pm$14    &  38$\pm$16   & 0.69  &                            & 1.01  &     \\
84 i,o     & 73  & 16 33 11.89 & -44 56 13.9    & 85$\pm$26    & 116$\pm$24   & 0.24  & 2MA 16331205-4456141 (1.7) & 1.89* & 1,3,5,Sz137  \\
85 i,o     & ... & 16 33 13.56 & -44 55 39.9    & 56$\pm$11    &  70$\pm$19   & 0.00  & 2MA 16331356-4455407 (0.8) & 0.95  & 1    \\
86 o       & 74  & 16 33 15.55 & -44 58 31.5    & 262$\pm$23   &  162$\pm$27  & 0.44  & 2MA 16331539-4458280 (3.8) & ...   & 6   \\
87         & ... & 16 33 15.62 & -44 50 58.3    & 54$\pm$17    &  100$\pm$24  & 0.60  & 2MA 16331568-4450568 (1.6) & 1.48  & 4,5     \\
88         & ... & 16 33 16.54 & -45 01 14.1    & 47$\pm$13    &  51$\pm$18   & 0.28  & ...                        & ...   &     \\
89         & ... & 16 33 16.76 & -45 00 53.1    & 32$\pm$10    &  28$\pm$14   & 0.02  & 2MA 16331662-4500539 (1.6) & 1.62* & 3,4,6 \\
90         & ... & 16 33 17.40 & -44 49 36.5    & 39$\pm$17    &  47$\pm$17   & 0.46  & 2MA 16331714-4449359 (2.7) & 2.69* & 5    \\
91 o       & ... & 16 33 17.50 & -44 56 04.6    & 42$\pm$10    &   8$\pm$10   & 0.60  &                            & ...   &     \\
92 o       & ... & 16 33 22.93 & -44 53 12.1    & 44$\pm$13    &  34$\pm$17   & 0.79  &                            & 2.78* &     \\
\enddata
\tablecomments{
In column 1, f = foreground source (Gaia parallax distance d $<$ 500 pc),
i = possible in-cloud source - there is at least one Gaia DR3 object within 3$''$
at d = 500 - 1000 pc,
o = on-cloud source, X-ray position is projected onto the darkest part of the cloud (Fig. 1).
X-ray positions (ra\_corr, dec\_corr header keywords) and  broad-band net 
counts (0.2-12 keV) are from the XMM-Newton pipeline processing source list. 
H.R.(pn) = pn\_net\_rate(2-12 keV)/pn\_net\_rate(0.5-12 keV). 
The relative position 
astrometry within each EPIC camera after applying corrections in spacecraft pointing
is accurate to $\approx$1$''$.5 (Kirsch et al. 2004). The mean uncertainty in XMM-Newton pipeline
source positions for the Norma detections is 1$''$.42$\pm$0$''$.94.
Possible 2MASS nIR counterparts lying within 3$''$ of the X-ray positions are identified as 2MA with the
XMM-2MASS position offset (arcsec) in parentheses after 2MASS Id.
VVV sources may also be present within 3$''$ of the X-ray positions but are only
listed (with position offsets in parentheses) if they have a nIR excess and no candidate 2MASS 
counterpart was found.
If one or more Gaia DR3 objects lie within 3$''$ of the X-ray position, 
the offset of the Gaia object lying closest to the X-ray position is given.
An asterisk follows if the offset is greater than the relative EPIC astrometric
uncertainty. 
\\
Notes. \\
       (1) There is a HST GSC class=0 (star) object within 3$''$ of the X-ray position.\\
       (2) There is a HST GSC class=3 (non-star) object within 3$''$ of the XMM-Newton position.\\
       (3) There is a a WISE ALL-Sky Survey IR object within 3$''$ of the XMM-Newton position. \\
       (4) There is more than one Gaia DR3 object within 3$''$ of the XMM-Newton position. \\
       (5) Source overlaps a pn CCD gap. Net pn count estimates and positions may be degraded. X-ray position 
           uncertainties are larger for the gap sources XMM 8 ($\pm$4$''$.0) and
           XMM 33 ($\pm$8$''$.2).    \\
       (6) The 2MASS identification is uncertain due to multiple 2MASS sources near the X-ray position
           or a large ($>$3$''$) position offset. \\
       (7) The identified VVV source has a nIR color excess. 
}
\end{deluxetable}

\clearpage
\begin{deluxetable}{lllccccccl}
\tablewidth{0pt}
\tablecaption{Summary of X-ray Spectral Fits}
\tablehead{
           \colhead{Object}               &
           \colhead{Alt. Id}              &
           \colhead{Model}                &
           \colhead{N$_{\rm H}$}          &
           \colhead{kT$_{1}$}             &
           \colhead{kT$_{2}$}             &
           \colhead{n$_{2}$/n$_{1}$}      &
           \colhead{$\Gamma_{ph}$}        &
           \colhead{$\chi^{2}$/dof}       &
           \colhead{F$_{x}$\tablenotemark{a}} \\
           \colhead{nr.}                  &
           \colhead{}                     &
           \colhead{}                     &
           \colhead{(10$^{22}$ cm$^{-2}$)}          &
           \colhead{(keV)}                &
           \colhead{(keV)}                &
           \colhead{}                     &
           \colhead{}                     &
           \colhead{}                     &
           \colhead{0.5 - 7 keV}                     
}
\startdata
XMM 78                  & CXO 64, Sz 136    & 1T apec & 1.10 [0.91-1.25] & 0.85$\pm$0.24 & ...       & ...   & ...         & 22.0/17 (1.30) & 0.96 (8.51)  \\
XMM 78                  & CXO 64, Sz 136    & 2T apec & 0.77 [0.32-1.08] & 0.84$\pm$0.24 & $>$3.2    & 0.95  & ...         & 12.9/15 (0.86) & 1.75 (4.95)\tablenotemark{b}  \\
\hline
XMM 84                  & CXO 73, Sz 137    & 1T apec & 1.11 [0.37-1.43] & 1.18$\pm$0.30 & ...       & ...   & ...         & 20.2/18 (1.12) & 0.96 (4.65)  \\
XMM 84                  & CXO 73, Sz 137    & 2T apec & 0.90 [0.43-1.22] & 1.08$\pm$0.30 & $>$3.2    & 0.52  & ...         & 17.3/16 (1.08) & 1.36 (3.92)\tablenotemark{c}  \\
\hline
CXO 4\tablenotemark{d}  & XMM 16            & 1T apec & 1.05 [0.83-1.45] & $>$5          & ...       & ...   & ...         & 39.8/41 (0.97) & 29.5 (44.1)  \\
CXO 4\tablenotemark{e}  & XMM 16            & 1T apec & 1.52 [1.06-2.03] & 2.30$\pm$0.60 & ...       & ...   & ...         & 10.6/10 (1.06) & 7.97 (20.3)  \\
XMM 16                  & CXO 4             & 1T apec & 1.41 [1.21-1.62]   & 2.27$\pm$0.30 & ...       & ...   & ...         & 49.8/46 (1.08) & 7.87 (19.6)  \\
XMM 16                  & CXO 4        & 1T apec$+$pl & 1.38 [1.16-1.65]   & [0.5]\tablenotemark{f} & ... & 38.3  & 2.6$\pm$0.3 & 38.4/45 (0.85) & 7.83 (26.5)  \\
\hline
CXO 21\tablenotemark{g}   & XMM 34            & 1T apec & 1.08             & $>$9            & ...       & ...   & ...         & 25.3/30 (0.84) & 11.5 (16.9)  \\
XMM 34                  & CXO 21            & 1T apec & 1.08 [0.81-1.38] & $>$13           & ...       & ...   & ...         & 33.2/31 (1.07) & 7.99 (11.5)  \\
XMM 34                  & CXO 21       & 1T apec$+$pl & 1.21 [0.78-1.72] & [0.5]\tablenotemark{f} & ...       & 3.8   & 1.4$\pm$0.3 & 33.2/30 (1.11) & 7.90 (13.4)  \\
\enddata
\tablecomments{
Notes: 
Object numbers refer to Tables 1 and 2. Chandra spectra were binned to a minimum of 10 cts/bin 
and XMM-Newton spectra to 10-20 cts/bin for fitting. Results for XMM-Newton sources are from 
fits of EPIC pn spectra except XMM 84 which is from simultaneous fits of MOS1 and MOS2 spectra.
One-temperature (1T apec) and two-temperature (2T apec) thermal plasma models are of the XSPEC form
N$_{\rm H}$*[kT$_{1}$ ($+$ kT$_{2}$)] with the absorption column density N$_{\rm H}$ determined using
the tbabs model. The XSPEC power-law (pl) model is of the form 
$K$exp($-$$\Gamma_{ph}$), $K$ has units photons/keV/cm$^{2}$/s at 1 keV, and
$\Gamma_{ph}$ (dimensionless) is the photon power-law index.
The value of n$_{2}$/n$_{1}$ is the ratio of XSPEC normalizations of hot/cool or pl/cool plasma components. 
Confidence ranges (in brackets), uncertainties, and kT lower limits are 1$\sigma$.
Temperatures (kT) are in energy units.
}
\tablenotetext{a}{The X-ray flux F$_{x}$ is the absorbed value followed in parentheses by 
the unabsorbed value.
Flux units are 10$^{-14}$ ergs cm$^{-2}$ s$^{-1}$.}
\tablenotetext{b}{The unabsorbed flux gives log L$_{x}$ = 30.64 ergs s$^{-1}$ at the Gaia DR3 parallax distance of 861.5 pc.}
\tablenotetext{c}{The unabsorbed flux gives log L$_{x}$ = 30.56 ergs s$^{-1}$ at the Gaia DR3 parallax distance of 879.5 pc.}
\tablenotetext{d}{Simultaneous fit of high-state spectra (ObsIds 24511, 27464).}
\tablenotetext{e}{Simultaneous fit of low-state spectra (ObsIds 23387, 26283).} 
\tablenotetext{f}{The value of kT was held fixed since the model does not converge to a stable value.} 
\tablenotetext{g}{Simultaneous fit of all four Chandra spectra with N$_{\rm H}$ held fixed at the value
                  determined from the 1T apec fit of the XMM 34 spectrum.}
\end{deluxetable}

\end{document}